\documentstyle[amsfonts,amssymb,aps,twocolumn]{revtex}
\input{epsf}

\def\lsim{\mathrel{\raise.3ex\hbox{$<$\kern-.75em\lower1ex\hbox{$\sim$}}}}
\def\gsim{\mathrel{\raise.3ex\hbox{$>$\kern-.75em\lower1ex\hbox{$\sim$}}}}
\newcommand{\ket}[1]{|{#1}\rangle}

\title{Topological quantum memory\thanks{CALT-68-2346}}
\author{Eric Dennis,$^{(1)}$\thanks{\tt edennis@princeton.edu} Alexei
Kitaev,$^{(2)}$\thanks{\tt kitaev@iqi.caltech.edu} Andrew Landahl,$^{(2)}$\thanks{\tt
alandahl@theory.caltech.edu} and John Preskill$^{(2)}$\thanks{\tt
preskill@theory.caltech.edu}}

\address{$^{(1)}$Princeton University, Princeton, NJ 08544, USA\\
$^{(2)}$Institute for Quantum Information, California Institute of Technology, 
Pasadena, CA 91125, USA }

\begin{document}

\maketitle

\begin{abstract}
We analyze {\em surface codes}, the topological quantum error-correcting codes introduced 
by Kitaev. In these codes, qubits are arranged in a two-dimensional array on a surface of nontrivial topology, and encoded quantum operations are associated with nontrivial homology cycles of the surface. We formulate protocols for error recovery, and study the efficacy of these protocols. An order-disorder phase transition occurs in this system at a nonzero critical value of the error rate; if the error rate is below the critical value (the {\em accuracy threshold}), encoded information can be protected arbitrarily well in the limit of a large code block. This phase transition can be accurately modeled by a three-dimensional $Z_2$ lattice gauge theory with quenched disorder. We estimate the accuracy threshold, assuming that all quantum gates are {\em local}, that qubits can be measured rapidly, and that polynomial-size classical computations can be executed instantaneously. We also devise a robust recovery procedure that does not require measurement or fast classical processing; however for this procedure the quantum gates are local only if the qubits are arranged in {\em four} or more spatial dimensions. We discuss procedures for encoding, measurement, and performing fault-tolerant universal quantum computation with surface codes, and argue that these codes provide a promising framework for quantum computing architectures.
\end{abstract}

\section{Introduction}
The microscopic world is quantum mechanical, but the macroscopic world is classical. This fundamental dichotomy arises because a coherent quantum superposition of two readily distinguishable macroscopic states is highly unstable. The quantum state of a macroscopic system rapidly {\em decoheres} due to unavoidable interactions between the system and its surroundings. 

Decoherence is so pervasive that it might seem to preclude subtle quantum interference phenomena in systems with many degrees of freedom. However, recent advances in the theory of quantum error correction suggest otherwise \cite{shor_9,steane_7}. We have learned that quantum states can be cleverly encoded so that the debilitating effects of decoherence, if not too severe, can be resisted. Furthermore, fault-tolerant protocols have been devised that allow an encoded quantum state to be reliably processed by a quantum computer with imperfect components \cite{shor_ft}. In principle, then, very intricate quantum systems can be stabilized and accurately controlled.

The theory of quantum fault tolerance has shown that, even for delicate coherent quantum states, information {\em processing} can prevent information {\em loss}. In this paper, we will study a particular approach to quantum fault tolerance that has notable advantages: in this approach, based on the {\em surface codes} introduced in \cite{kitaev1,kitaev2}, the quantum processing needed to control errors has especially nice locality properties. For this reason, we think that surface codes suggest a particularly promising approach to quantum computing architecture.

One glittering achievement of the theory of quantum fault tolerance is the {\em threshold theorem}, which asserts that an arbitrarily long quantum computation can be executed with arbitrarily high reliability, provided that the error rates of the computer's fundamental quantum gates are below a certain critical value, the {\em accuracy threshold} \cite{knill,ben-or,kitaev_threshold,jp_threshold,gottesman_threshold}. The numerical value of this accuracy threshold is of great interest for future quantum technologies, as it defines a standard that should be met by designers of quantum hardware. The critical error probability per gate $p_c$ has been estimated as $p_c\gsim 10^{-4}$; very roughly speaking, this means that robust quantum computation is possible if the decoherence time of stored qubits is at least $10^4$ times longer than the time needed to execute one fundamental quantum gate \cite{jp_gott}, assuming that decoherence is the only source of error.

This estimate of the accuracy threshold is obtained by analyzing the efficacy of a {\em concatenated code}, a hierarchy of codes within codes, and it is based on many assumptions, which we will elaborate in Sec.~\ref{sec:ft}. For now, we just emphasize one of these assumptions: that a quantum gate can act on any pair of qubits, with a fidelity that is independent of the spatial separation of the qubits. This assumption is clearly unrealistic; it is made because it greatly simplifies the analysis. Thus this estimate will be reasonable for a practical device only to the extent that the hardware designer is successful in arranging that qubits that must interact are kept close to one another. It is known that the threshold theorem still applies if quantum gates are required to be local \cite{ben-or,gottesman_local}, but for this realistic case careful estimates of the threshold have not been carried out.

We will perform a quite different estimate of the accuracy threshold, based on surface codes rather than concatenated codes. This estimate applies to a device with strictly local quantum gates, if the device is controlled by a classical computer that is perfectly reliable, and whose clock speed is much faster than the clock speed of the quantum computer. In this approach, some spatial nonlocality in effect is still allowed, but we demand that all the nonlocal processing be classical. Specifically, an error syndrome is extracted by performing local quantum gates and measurements; then a classical computation is executed to infer what quantum gates are needed to recover from error. We will assume that this classical computation, which actually requires a time bounded above by a polynomial in the number of qubits in the quantum computer, can be executed in a constant number of time steps. Under this assumption, the existence of an accuracy threshold can be established and its value can be estimated. If we assume that the classical computation can be completed in a single time step, we estimate that the critical error probability $p_c$  per qubit and per time step satisfies $p_c\ge 1.7\times 10^{-4}$. This estimate applies to the accuracy threshold for reliable {\em storage} of quantum information, rather than for reliable processing. The threshold for quantum computation is not as easy to analyze definitively, but we will argue that its numerical value is not likely to be substantially different.

We believe that principles of fault tolerance will dictate the shape of future quantum computing architectures. In  Sec.~\ref{sec:ft} we compile a list of hardware features that are conducive to fault-tolerant processing, and outline the design of a fault-tolerant quantum computer that incorporates surface coding. We review the properties of surface codes in Sec.~\ref{sec:surface}, emphasizing in particular that the qubits in the code block can be arranged in a {\em planar} sheet \cite{brav_kit,freed_meyer}, and that errors in the syndrome measurement complicate the recovery procedure.  The core of the paper is Sec.~\ref{sec:statphys}, where we relate recovery from errors using surface codes to a statistical-mechanical model with local interactions. In the (unrealistic) case where syndrome measurements are perfect, this model becomes the two-dimensional Ising model with quenched disorder, whose phase diagram has been studied by Monte Carlo simulations. These simulations indicate that if the syndrome information is put to optimal use, error recovery succeeds with a probability that approaches one in the limit of a large code block, if and only if both phase errors and bit-flip errors occur with a probability per qubit less than about $11\%$. In the more realistic case where syndrome measurements are imperfect, error recovery is modeled by a three-dimensional $Z_2$ gauge theory with quenched disorder, whose phase diagram (to the best of our knowledge) has not been studied previously. The third dimension that arises can be interpreted as time --- since the syndrome information cannot be trusted, we must repeat the measurement many times before we can be confident about the correct way to recover from the errors. We argue that an order-disorder phase transition of this model corresponds to the accuracy threshold for quantum storage, and furthermore that the optimal recovery procedure can be computed efficiently on a classical computer. We proceed in Sec.~\ref{sec:chains_minimal} to prove a rather crude lower bound on the accuracy threshold, concluding that error recovery procedure is sure to succeed in the limit of a large code block under suitable conditions: for example, if in each round of syndrome measurement, qubit phase errors, qubit bit-flip errors, and syndrome bit errors all occur with probability below $1.14\%$. Tighter estimates of the accuracy threshold could be obtained through numerical studies of the quenched gauge theory. 

In deriving this accuracy threshold for quantum storage, we assumed that an unlimited amount of syndrome data could be deposited in a classical memory, if necessary. But in Sec.~\ref{sec:finite_time} we show that this threshold, and a corresponding accuracy threshold for quantum computation, remain intact even if the classical memory is limited to polynomial size. Then in Sec.~\ref{sec:storage_gates} we analyze quantum circuits for syndrome measurement, so that our estimate of the accuracy threshold can be reexpressed as a fidelity requirement for elementary quantum gates. We conclude that our quantum memory can resist decoherence if gates can be executed in parallel, and if the qubit decoherence time is at least $6000$ times longer than the time needed to execute a gate. In Sec.~\ref{sec:enc_meas} we show that encoded qubits can be accurately prepared and reliably measured. We also describe how a surface code with a small block size can be built up gradually to a large block size; this procedure allows us to enter a qubit in an unknown quantum state into our quantum memory with reasonable fidelity, and then to maintain that fidelity for an indefinitely long time. We explain in Sec.~\ref{sec:ftgates} how a universal set of quantum gates acting on protected quantum information can be executed fault-tolerantly.

Most of the analysis of the accuracy threshold in this paper is premised on the assumption that qubits can be measured quickly and that classical computations can be done instantaneously and perfectly. In Sec.~\ref{sec:four_dim} we drop these assumptions. We devise a recovery procedure that does not require measurement or classical computation, and infer a lower bound on the accuracy threshold. Unfortunately, though, the quantum processing in our procedure is not spatially local unless the dimensionality of space is at least four. Sec.~\ref{sec:conclusions} contains some concluding remarks.

This paper analyzes applications of surface coding to quantum memory and quantum computation that could in principle be realized in any quantum computer that meets the criteria of our computational model, whatever the details of how the local quantum gates are physically implemented. It has also been emphasized \cite{kitaev1,kitaev2} that surface codes may point the way toward realizations of intrinsically stable quantum memories ({\em physical} fault tolerance). In that case, protection against decoherence would be achieved without the need for active information processing, and how accurately the protected quantum states can be processed might depend heavily on the details of the implementation.

\section{Fault tolerance and quantum architecture}
\label{sec:ft}

To prove that a quantum computer with noisy gates can perform a robust quantum computation, we must make some assumptions about the nature of the noise and about how the computer operates. In fact, similar assumptions are needed to prove that a classical computer with noisy gates is robust \cite{gacs}. Still, it is useful to list these requirements --- they should always be kept in mind when we contemplate proposed schemes for building quantum computing hardware:

\begin{itemize}
\item {\em Constant error rate}. We assume that the strength of the noise is independent of the number of qubits in the computer. If the noise increases as we add qubits, then we cannot reduce the error rate to an arbitrarily low value by increasing the size of the code block. 
\item {\em Weakly correlated errors}. Errors must not be too strongly correlated, either in space or in time. In particular, fault-tolerant procedures fail if errors act simultaneously on many qubits in the same code block. If possible, the hardware designer should strive to keep qubits in the same block isolated from one another.
\item {\em Parallel operation}. We need to be able to perform many quantum gates in a single time step. Errors occur at a constant rate per unit time, and we are to control these errors through information processing. We could never keep up with the accumulating errors except by doing processing in different parts of the computer at the same time.
\item {\em Reusable memory}. Errors introduce entropy into the computer, which must be flushed out by the error recovery procedure. Quantum processing transfers the entropy from the qubits that encode the protected data to ``ancilla'' qubits that can be discarded. Thus fresh ancilla qubits must be continually available. The ability to erase (or replace) the ancilla quickly is an essential hardware requirement \cite{nisan}.
\end{itemize}

In some estimates of the threshold, additional assumptions are made. While not strictly necessary to ensure the existence of a threshold, these assumptions may be useful, either because they simplify the analysis of the threshold or because they allow us to increase its numerical value. Hence these assumptions, too, should command the attention of the prospective hardware designer:

\begin{itemize}
\item {\em Fast measurements}. It is helpful to assume that a qubit can be measured as quickly as a quantum gate can be executed. For some implementations, this may not be a realistic assumption --- measurement requires the amplification of a microscopic quantum effect to a macroscopic signal, which may take a while. But by measuring a classical error syndrome for each code block, we can improve the efficiency of error recovery. Furthermore, if we can measure qubits and perform quantum gates conditioned on classical measurement outcomes, then we can erase ancilla qubits by projecting onto the $\{|0\rangle,|1\rangle\}$ basis and flipping the qubit if the outcome is $|1\rangle$.
\item {\em Fast and accurate classical processing}. If classical processing is faster and more accurate than quantum processing, then it is beneficial to substitute classical processing for quantum processing when possible. In particular, if the syndrome is measured, then a classical computation can be executed to determine how recovery should proceed. Ideally, the classical processors that coordinate the control of the quantum computer should be integrated into the quantum hardware.
\item {\em No leakage.} It is typically assumed that, though errors may damage the state of the computer, the qubits themselves remain accessible --- they do not ``leak'' out of the device. In fact, at least some types of leakage can be readily detected. If leaked qubits, once detected, can be replaced easily by fresh qubits, then leakage need not badly compromise performance. Hence, a desirable feature of hardware is that leaks are easy to detect and correct.
\item {\em Nonlocal quantum gates}. Higher error rates can be tolerated, and the estimate of the threshold is simplified, if we assume that two-qubit quantum gates can act on any pair of qubits with a fidelity independent of the distance between the qubits. However useful, this assumption is not physically realistic. What the hardware designer can and should do, though, is try to arrange that qubits that will need to interact with one another are kept close to one another. In particular, the ancilla qubits that absorb entropy should be carefully integrated into the design \cite{gottesman_local}.
\end{itemize}

If we do insist that all quantum gates are local, then another desirable feature is:

\begin{itemize}
\item {\em High coordination number}. A threshold theorem applies even if qubits form a one-dimensional array \cite{ben-or,gottesman_local}. But local gates are more effective if the qubits are arranged in three dimensions, so that each qubit has more neighbors.
\end{itemize}

Suppose, then, that we are blessed with an implementation of quantum computation that meets all of our desiderata. Qubits are arranged in a three-dimensional lattice, and can be projectively measured quickly. Reasonably accurate quantum gates can be applied in parallel to single qubits or to neighboring pairs of qubits. Fast classical processing is integrated into the qubit array. Under these conditions planar surface codes provide an especially attractive way to operate the quantum computer fault-tolerantly.

We may envision our quantum computer as a stack of planar sheets, with a protected logical qubit encoded in each sheet. Adjacent to each logical sheet is an associated sheet of ancilla qubits that are used to measure the error syndrome of that code block; after each measurement, these ancilla qubits are erased and then immediately reused. Encoded two-qubit gates can be performed between neighboring logical sheets, and any two logical sheets in the stack can be brought into contact by performing swap gates that move the sheets through the intervening layers of logical and ancilla qubits. As a quantum circuit is executed in the stack, error correction is continually applied to each logical sheet to protect against decoherence and other errors. Portions of the stack are designated as ``software factories,'' where special ancilla states are prepared and purified --- this software is then consumed during the execution of certain quantum gates that cannot be implemented directly. 

A notable feature of this design (or other fault-tolerant designs) is that most of the information processing in the device is devoted to controlling errors, rather than moving the computation forward. How accurately must the fundamental quantum gates be executed for this error control to be effective, so that our machine is computationally powerful? Our goal in this paper is to address this question.

\section{Surface codes}
\label{sec:surface}

We will study the family of quantum error-correcting codes introduced in \cite{kitaev1,kitaev2}.
These codes are especially well suited for fault-tolerant implementation, because the procedure for measuring the error syndrome is highly local. 

\subsection{Toric codes}
For the code originally described in \cite{kitaev1,kitaev2}, it is convenient to imagine that the qubits are in one-to-one correspondence with the links of a square lattice drawn on a torus, or, equivalently, drawn on a square with opposite edges identified. Hence we will refer to them as ``toric codes.''  Toric codes can be generalized to a broader class of quantum codes, with each code in the class associated with a tessellation of a two-dimensional surface. Codes in this broader class will be called ``surface codes.''

A surface code is a special type of ``stabilizer code'' \cite{att,gott_stab}. A (binary) stabilizer code can be characterized as the simultaneous eigenspace with eigenvalue one of a set of mutually commuting check operators (or ``stabilizer generators''), where each generator is a ``Pauli operator.'' We use the notation
\begin{eqnarray}
& &I= \pmatrix{1 & 0\cr 0 & 1\cr}~,~\quad X=\pmatrix{0 & 1\cr 1 & 0\cr}~,\\
& &Y= \pmatrix{0 & -i\cr i & 0\cr}~, ~Z=\pmatrix{1 & 0\cr 0 & -1\cr}
\end{eqnarray}
for the $2\times 2$ identity and Pauli matrices; a Pauli operator acting on $n$ qubits is one of the $2^{2n}$ tensor product operators
\begin{equation}
\{I,X,Y,Z\}^{\otimes n}~.
\end{equation}

\begin{figure}
\begin{center}
\leavevmode
\epsfxsize=3in
\epsfbox{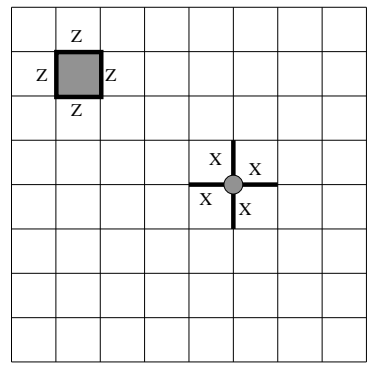}
\end{center}
\caption{Check operators of the toric code. Each plaquette operator is a tensor product of $Z$'s acting on the four links contained in the plaquette. Each site operator is a tensor product of $X$'s acting on the four links that meet at the site.}
\label{fig_checks}
\end{figure}

For the toric code defined by the $L\times L$ square lattice on the torus, there are $2L^2$ links of the lattice, and hence $2L^2$ qubits in the code block. Check operators are associated with each site and with each elementary cell (or ``plaquette'') of the lattice, as shown in Fig.~\ref{fig_checks}. The check operator at site $s$ acts nontrivially on the four links that meet at the site; it is the tensor product 
\begin{equation}
X_s=\otimes_{\ell\ni s} X_\ell
\end{equation}
acting on those four qubits, times the identity acting on the remaining qubits. The check operator at plaquette $P$ acts nontrivially on the four links contained in the plaquette, as the tensor product 
\begin{equation}
Z_P=\otimes_{\ell\in P} Z_\ell~,
\end{equation}
times the identity on the remaining links.

Although $X$ and $Z$ anticommute, the check operators are mutually commuting. Obviously, site operators commute with site operators, and plaquette operators with plaquette operators. Site operators commute with plaquette operators because a site operator and a plaquette operator act either on disjoint sets of links, or on sets whose intersection contains two links. In the former case, the operators obviously commute, and in the latter case, two canceling minus signs arise when the site operator commutes through the plaquette operator. The check operators generate an Abelian group, the code's stabilizer.

The check operators can be simultaneously diagonalized, and the toric code is the space in which each check operator acts trivially.  Because of the periodic boundary conditions, each site or plaquette operator can be expressed as the product of the other $L^2-1$ such operators; the product of all $L^2$ site operators or all $L^2$ plaquette operators is the identity, since each link operator occurs twice in the product, and $X^2=Z^2=I$. There are no further relations among these operators; therefore, there are $2\cdot (L^2-1)$ independent check operators, and hence two encoded qubits (the code subspace is four dimensional).

A Pauli operator that commutes with all the check operators will preserve the code subspace.  What operators have this property? To formulate the answer, it is convenient to recall some standard mathematical terminology. A mapping that assigns an element of $Z_2=\{0,1\}$ to each link of the lattice is called a ($Z_2$-valued) {\em 1-chain}. In a harmless abuse of language, we will also use the term 1-chain (or simply chain) to refer to the set of all links that are assigned the value 1 by such a mapping. The 1-chains form a vector space over $Z_2$ --- intuitively, the sum $u+v$ of two chains $u$ and $v$ is a disjoint union of the links contained in the two 1-chains. Similarly, 0-chains assign elements of $Z_2$ to lattice sites and 2-chains assign elements of $Z_2$ to lattice plaquettes; these also form vector spaces. A linear boundary operator $\partial$ can be defined that takes 2-chains to 1-chains and 1-chains to 0-chains: the boundary of a plaquette is the sum of the four links comprising the plaquette, and the boundary of a link is the sum of the two sites at the ends of the link. A chain whose boundary is trivial is called a {\em cycle}.

Now, any Pauli operator can be expressed as a tensor product of $X$'s (and $I$'s) times a tensor product of $Z$'s (and $I$'s).  The tensor product of $Z$'s and $I$'s defines a $Z_2$-valued 1-chain, where links acted on by $Z$ are mapped to 1 and links acted on by $I$ are mapped to 0. This operator trivially commutes with all of the plaquette check operators, but commutes with a site operator if and only if an even number of $Z$'s act on the links adjacent to the site.  Thus, the corresponding 1-chain must be a cycle. Similarly, the tensor product of $X$'s trivially commutes with the site operators, but commutes with a plaquette operator only if an even number of $X$'s act on the links contained in the plaquette. This condition can be more conveniently expressed if we consider the dual lattice, in which sites and plaquettes are interchanged; the links dual to those on which $X$ acts form a cycle of the dual lattice.  In general, then, a Pauli operator that commutes with the stabilizer of the code can be represented as a tensor product of $Z$'s acting on a cycle of the lattice, times a tensor product of $X$'s acting on a cycle of the dual lattice.

\begin{figure}
\begin{center}
\leavevmode
\epsfxsize=3in
\epsfbox{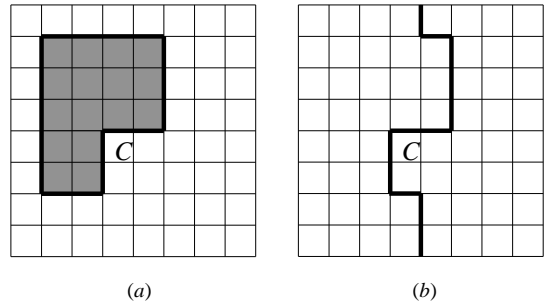}
\end{center}
\caption{Cycles on the lattice. $(a)$ A homologically trivial cycle bounds a region that can be tiled by plaquettes. The corresponding tensor product of $Z$'s lies in the stabilizer of the toric code. $(b)$ A homologically nontrivial cycle is not a boundary. The corresponding tensor product of $Z$'s commutes with the stabilizer but is not contained in it. It is a logical operation that acts nontrivially in the code subspace.}
\label{fig_homo}
\end{figure}

Cycles are of two distinct types.  A 1-cycle is {\em homologically trivial} if it can be expressed as the boundary of a 2-chain (Fig.~\ref{fig_homo}a).  Thus, a homologically trivial cycle on our square lattice has an interior that can be ``tiled'' by plaquettes, and a product of $Z$'s acting on the links of the cycle can be expressed as a product of the enclosed plaquette operators.  This operator is therefore a product of the check operators --- it is contained in the code stabilizer and acts trivially on the code subspace. Similarly, a product of $X$'s acting on links that comprise a homologically trivial cycle of the dual lattice is also a product of check operators. Furthermore, {\em any} element of the stabilizer group of the toric code (any product of the generators) can be expressed as a product of $Z$'s acting on a homologically trivial cycle of the lattice times $X$'s acting on a homologically trivial cycle of the dual lattice.

\begin{figure}
\begin{center}
\leavevmode
\epsfxsize=3in
\epsfbox{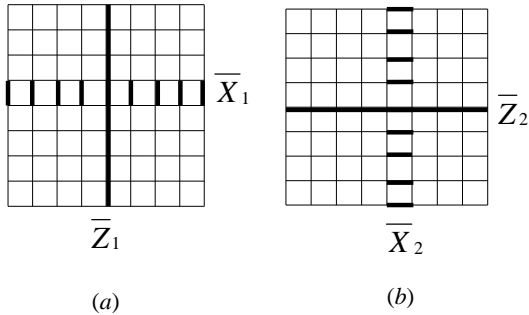}
\end{center}
\caption{Basis for the operators that act on the two encoded qubits of the toric code. The logical operators $\bar Z_1$ and $\bar Z_2$ are tensor products of $Z$'s associated with the fundamental nontrivial cycles of the torus constructed from links of the lattice. The complementary operators $\bar X_1$ and $\bar X_2$ are tensor products of $X$'s associated with  nontrivial cycles constructed from links of the dual lattice.}
\label{fig_logical}
\end{figure}

But a cycle could be homologically nontrivial, that is, not the boundary of anything (Fig.~\ref{fig_homo}b). A product of $Z$'s corresponding to a nontrivial cycle commutes with the code stabilizer (because it is a cycle), but is not contained in the stabilizer (because the cycle is nontrivial). Therefore, while this operator preserves the code subspace, it acts nontrivially on encoded quantum information. Associated with the two fundamental nontrivial cycles of the torus, then, are the encoded operations $\bar Z_1$ and $\bar Z_2$ acting on the two encoded qubits.  Associated with the two dual cycles of the dual lattice are the corresponding encoded operations $\bar X_1$ and $\bar X_2$, as shown in Fig~\ref{fig_logical}.

A Pauli operator acting on $n$ qubits is said to have {\em weight} $w$ if the identity $I$ acts on $n-w$ qubits and nontrivial Pauli matrices act on $w$ qubits. The {\em distance} $d$ of a stabilizer code is the weight of the minimal-weight Pauli operator that preserves the code subspace and acts nontrivially within the code subspace. If an encoded state is damaged by the action of a Pauli operator whose weight is less than half the code distance, then we can recover from the error successfully by applying the minimal weight Pauli operator that returns the damaged state to the code subspace (which can be determined by measuring the check operators). For a toric code, the distance is the number of lattice links contained in the shortest homologically nontrivial cycle on the lattice or dual lattice. Thus in the case of an $L\times L$ square lattice drawn on the torus, the code distance is $d=L$. 

The great virtue of the toric code is that the check operators are so simple. Measuring a check operator requires a quantum computation, but because each check operator involves just four qubits in the code block, and these qubits are situated near one another, the measurement can be executed by performing just a few quantum gates.  Furthermore, the ancilla qubits used in the measurement can be situated where they are needed, so that the gates act on pairs of qubits that are in close proximity.

The observed values of the check operators provide a ``syndrome'' that we may use to diagnose errors. If there are no errors in the code block, then every check operator takes the value 1. Since each check operator is associated with a definite position on the surface, a site of the lattice or the dual lattice, we may describe the syndrome by listing all positions where the check operators take the value $-1$. It is convenient to regard each such position as the location of a particle, a ``defect'' in the code block.

\begin{figure}
\begin{center}
\leavevmode
\epsfxsize=3in
\epsfbox{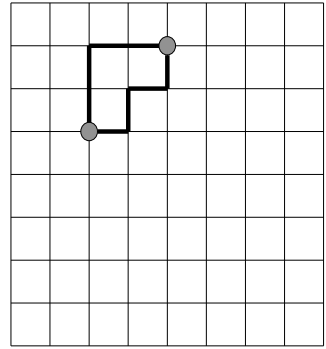}
\end{center}
\caption{The highly ambiguous syndrome of the toric code. The two site defects shown could arise from errors on either one of the two chains shown. In general, error chains with the same boundary generate the same syndrome, and error chains that are homologically equivalent act on the code space in the same way.}
\label{fig_defect}
\end{figure}

If errors occur on a particular chain (a set of links of the lattice or dual lattice), then defects occur at the sites on the {\em boundary} of the chain. Evidently, then, the syndrome is highly ambiguous, as many error chains can share the same boundary, and all generate the same syndrome. For example, the two chains shown in Fig.~\ref{fig_defect} end on the same two sites.  If errors occur on one of these chains, we might incorrectly infer that the errors actually occured on the other chain.  Fortunately, though, this ambiguity need not cause harm.  If $Z$ errors occur on a particular chain, then by applying $Z$ to each link of {\em any} chain with the same boundary as the actual error chain, we will successfully remove all defects. Furthermore, as long as the chosen chain is {\em homologically} correct (differs from the actual error chain by the one-dimensional boundary of a two-dimensional region), then the encoded state will be undamaged by the errors.  In that event, the product of the actual $Z$ errors and the $Z$'s that we apply is contained in the code stabilizer and therefore acts trivially on the code block. 

Heuristically, an error chain can be interpreted as a physical process in which a defect pair nucleates, and the two members of the pair drift apart. To recover from the errors, we lay down a ``recovery chain'' bounded by the two defect positions, which we can think of as a physical process in which the defects are brought together to reannihilate. If the defect world line consisting of both the error chain and the recovery chain is homologically trivial, then the encoded quantum state is undamaged. But if the world line is homologically nontrivial (if the two members of the pair wind around a cycle of the torus before reannihilating), then an error afflicts the encoded quantum state.

\subsection{Planar codes}

If all check operators are to be readily measured with local gates, then the qubits of the toric code need to be arranged on a topologically nontrivial surface, the torus, with the ancilla qubits needed for syndrome measurement arranged on an adjacent layer. In practice, the toroidal topology is likely to be inconvenient, especially if we want qubits residing in different tori to interact with one another in the course of a quantum computation. Fortunately, surface codes can be constructed in which all check operators are local and the qubits are arranged on planar sheets \cite{brav_kit,freed_meyer}. The planar topology will be more conducive to realistic quantum computing architectures.

\begin{figure}
\begin{center}
\leavevmode
\epsfxsize=3in
\epsfbox{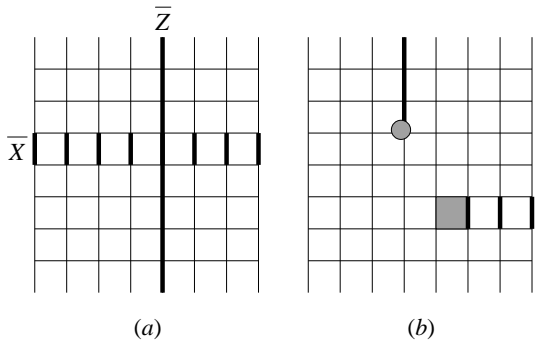}
\end{center}
\caption{A planar quantum code. $(a)$ At the top and bottom are the ``plaquette edges'' (or ``rough edges'') where there are three-qubit plaquette operators, and at the left and right are the ``site edges'' (or ``smooth edges'') where there are three-qubit site operators. The logical operation $\bar Z$ for the one encoded qubit is a tensor product of $Z$'s acting on a chain running from one rough edge to the other, and the logical operation $\bar X$ is a tensor product of $X$'s acting on a chain of the dual lattice running from one smooth edge to the other. For the lattice shown, the code's distance is $L=8$. $(b)$ Site and plaquette defects can appear singly, rather than in pairs. An isolated site defect arises from an error chain that ends at a rough edge, and an isolated plaquette defect arises from a dual error chain that ends at a smooth edge.} 
\label{fig_planar}
\end{figure}

In the planar version of the surface code, there is a distinction between the check operators at the boundary of the surface and the check operators in the interior.  Check operators in the interior are four-qubit site or plaquette operators, and those at the boundary are three-qubit operators.  Furthermore, the boundary has two different types of edges as shown in Fig.~\ref{fig_planar}. Along a ``plaquette edge''  or ``rough edge,'' each check operator is a three-qubit plaquette operator $Z^{\otimes 3}$. Along a ``site edge'' or ``smooth edge,'' each check operator is a three-qubit site operator $X^{\otimes 3}$. 

As before, in order to commute with the code stabilizer, a product of $Z$'s must act on an even number of links adjacent to each site of the lattice. Now, though, the links acted upon by $Z$'s may comprise an {\em open} path that begins and ends on a rough edge. We may then say that the 1-chain comprised of all links acted upon by $Z$ is a cycle {\em relative to the rough edges}. Similarly, a product of $X$'s that commutes with the stabilizer acts on a set of links of the dual lattice that comprise a cycle relative to the smooth edges.  

Cycles relative to the rough edges come in two varieties. If the chain contains an even number of the free links strung along the rough edge, then it can be tiled by plaquettes (including the boundary plaquettes), and so the corresponding product of $Z$'s is contained in the stabilizer. We say that the relative 1-cycle is a relative boundary of a 2-chain. However, a chain that stretches from one rough edge to another is not a relative boundary --- it is a representative of a nontrivial relative homology class. The corresponding product of $Z$'s commutes with the stabilizer but does not lie in it, and we may take it to be the logical operation $\bar Z$ acting on an encoded logical qubit.  Similarly, cycles relative to the smooth edges also come in two varieties, and a product of $X$'s associated with the nontrivial relative homology cycle of the dual lattice may be taken to be the logical operation $\bar X$ (see Fig.~\ref{fig_planar}a).

A code with distance $L$ is obtained from a square lattice, if the shortest paths from rough edge to rough edge, and from smooth edge to smooth edge, both contain $L$ links.  The lattice has $L^2 +(L-1)^2$ links,  $L(L-1)$ plaquettes, and $L(L-1)$ sites.  Now all plaquette and site operators are independent, which is another way to see that the number of encoded qubits is $L^2 +(L-1)^2- 2L(L-1)=1$. 

The distinction between a rough edge and a smooth edge can also be characterized by the behavior of the defects at the boundary, as shown in Fig.~\ref{fig_planar}b.  In the toric codes, defects always appear in pairs, because every 1-chain has an even number of boundary points. But for planar codes, individual defects can appear, since a 1-chain can terminate on a rough edge.  Thus a propagating site defect can reach the rough edge and disappear. But if the site defect reaches the smooth edge, it persists at the boundary. Similarly, a plaquette defect can disappear at the smooth edge, but not at the rough edge. 

Let us briefly note some generalizations of the toric codes and planar codes that we have described. First, there is no need to restrict attention to lattices that have coordination number 4 at each site and plaquette. Any tessellation of a surface (and its dual tessellation) can be associated with a quantum code. Second, we may consider surfaces of higher genus.  For a closed orientable Riemann surface of genus $g$, $2g$ qubits can be encoded --- each time a handle is added to the surface, there are two new homology cycles and hence two new logical $\bar Z$'s. The distance of the code is the length of the shortest nontrivial cycle on lattice or dual lattice. For planar codes, we may consider a surface with $e$ distinct rough edges separated by $e$ distinct smooth edges. Then $e-1$ qubits can be encoded, associated with the relative 1-cycles that connect one rough edge with any of the others. The distance is the length of the shortest path reaching from one rough edge to another, or from one smooth edge to another on the dual lattice. Alternatively, we can increase the number of encoded qubits stored in a planar sheet by punching holes in the lattice. For example, if the outer boundary of the surface is a smooth edge, and there are $h$ holes, each bounded by a smooth edge, then $h$ qubits are encoded. For each hole, a cycle on the lattice that encloses the hole is associated with the corresponding logical $\bar Z$, and a path on the dual lattice from the boundary of the hole to the outer boundary is associated with the logical $\bar X$.

If (say) phase errors are more common than bit-flip errors, quantum information can be stored more efficiently with an {\em asymmetric} planar code, such that the distance from rough edge to rough edge is longer than the distance from smooth edge to smooth edge. However, these asymmetric codes are less convenient for processing of the encoded information. 

The surface codes can also be generalized to higher dimensional manifolds, with logical operations again associated with homologically nontrivial cycles.  In Sec.~\ref{sec:four_dim}, we will discuss a four-dimensional example. 

\subsection{Fault-tolerant recovery}
\label{subsec:ftrec}
A toric code defined on a lattice of linear size $L$ has block size $2L^2$ and distance $L$. Therefore, if the probability of error per qubit is $p$, the number of errors expected in a large code block is of order $pL^2$, and therefore much larger than the code distance.

However, the performance of a toric code is much better than would be guessed naively based on its distance. In principle, $L/2$ errors could suffice to cause damage to the encoded information. But in fact this small number of errors can cause irrevocable damage only if the distribution of the errors is highly atypical.

\begin{figure}
\begin{center}
\leavevmode
\epsfxsize=3in
\epsfbox{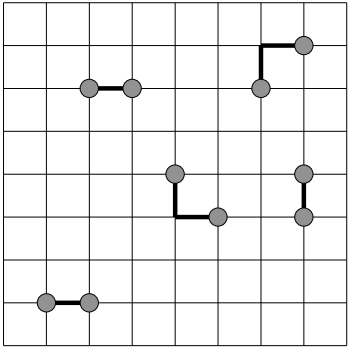}
\end{center}
\caption{Pairs of defects. If the error rate is small and errors on distinct links are uncorrelated, then connected error chains are typically short and the positions of defects are highly correlated. It is relatively easy to guess how the defects should be paired up so that each pair is the boundary of a connected chain.}
\label{fig_dilute}
\end{figure}

If the error probability $p$ is small, then links where errors occur (``error links'') are dilute on the lattice. Long connected chains of error links are quite rare, as indicated in Fig.~\ref{fig_dilute}. It is relatively easy to guess a way to pair up the observed defects that is homologically equivalent to the actual error chain. Hence we expect that a number of errors that scales {\em linearly} with the block size can be tolerated.  That is, if the error probability $p$ per link is small enough, we expect to be able to recover correctly with a probability that approaches one as the block size increases. We therefore anticipate that there is an accuracy threshold for storage of quantum information using a toric code.

\begin{figure}
\begin{center}
\leavevmode
\epsfxsize=3in
\epsfbox{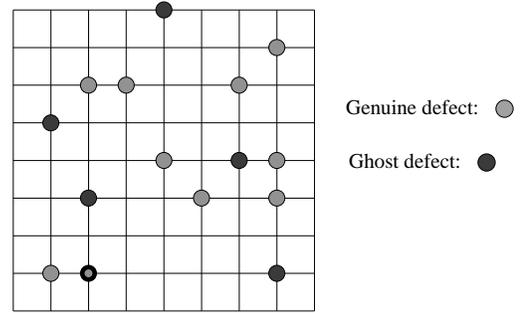}
\end{center}
\caption{Ghost defects. Since faults can occur in the measurement of the error syndrome, the measured syndrome includes both genuine defects (lightly shaded) associated with actual errors and phony ``ghost defects'' (darkly shaded) that arise at randomly distributed locations. To perform recovery successfully, we need to be able to distinguish reliably between the genuine defects and the ghost defects. The position that is shaded both lightly and darkly represents a genuine defect that goes unseen due to a measurement error.}
\label{fig_ghosts}
\end{figure}

Unfortunately, life is not quite so simple, because the measurement of the syndrome will not be perfect. Occasionally, a faulty measurement will indicate that a defect is present at a site even though no defect is actually there, and sometimes an actual defect will go unobserved. Hence the population of real defects (which have strongly correlated positions) will be obscured by a population of phony ``ghost defects'' and ``missing defects'' (which have randomly distributed positions), as in Fig.~\ref{fig_ghosts}.

Therefore, we should execute recovery cautiously. It would be dangerous to blithely proceed by flipping qubits on a chain of links bounded by the observed defect positions. Since a ghost defect is typically far from the nearest genuine defect, this procedure would introduce many additional errors --- what was formerly a ghost defect would become a real defect connected to another defect by a long error chain. Instead we must repeat the syndrome measurement an adequate number of times to verify its authenticity. It is subtle to formulate a robust recovery procedure that incorporates repeated measurements, since further errors accumulate as the measurements are repeated and the gas of defects continues to evolve.

We know of three general strategies that can be invoked to achieve robust macroscopic control of a system that is subjected to microscopic disorder. One method is to introduce a hierarchical organization in such a way that effects of noise get weaker and weaker at higher and higher levels of the hierarchy. This approach is used by G\'acs \cite{gacs} in his analysis of robust one-dimensional classical cellular automata, and also in concatenated quantum coding \cite{knill,ben-or,kitaev_threshold,jp_threshold,gottesman_threshold}. A second method is to introduce more spatial dimensions. A fundamental principle of statistical physics is that local systems with higher spatial dimensionality and hence higher coordination number are more resistant to the disordering effects of fluctuations.  In Sec.~\ref{sec:four_dim} we will follow this strategy in devising and analyzing a topological code that has nice locality properties in four dimensions. From the perspective of block coding, the advantage of extra dimensions is that local check operators can be constructed with a higher degree of redundancy, which makes it easier to reject faulty syndrome information.

In the bulk of this paper we will address the issue of achieving robustness through a third strategy, namely by introducing a modest amount of nonlocality into our recovery procedure. But we will insist that all quantum processing is strictly local; the nonlocality will be isolated in {\em classical} processing. Specifically, to decide on the appropriate recovery step, a classical computation will be performed whose input is an error syndrome measured at all the sites of the lattice. We will require that this classical computation can be executed in a time bounded by a polynomial in the number of lattice sites. For the purpose of estimating the accuracy threshold, we will imagine that the classical calculation is instantaneous and perfectly accurate. 

Our approach is guided by the expectation that quantum computers will be slow and unreliable while classical computers are fast and accurate. It is advantageous to replace quantum processing by classical processing if the classical processing can accomplish the same task.

\subsection{Surface codes and physical fault tolerance}
In this paper, we regard the surface codes as block quantum error-correcting codes with properties that make them especially amenable to fault-tolerant quantum storage and computation. But we also remark here that because of the locality of the check operators, these codes admit another tempting interpretation that was emphasized in \cite{kitaev1,kitaev2}.

Consider a model physical system, with qubits arranged in a square lattice, and with a (local) Hamiltonian that can be expressed as minus the sum of the check operators of a surface code. Since the check operators are mutually commuting, we can diagonalize the Hamiltonian by diagonalizing each check operator separately, and its degenerate ground state is the code subspace. Thus, a real system that is described well enough by this model could serve as a robust quantum memory.  

The model system has several crucial properties. First of all, it has a mass gap, so that its qualitative properties are stable with respect to generic weak local perturbations. Secondly, it has two types of localized quasiparticle excitations, the site defects and plaquette defects. And third, there is an exotic long-range interaction between a site defect and a plaquette defect. 

The interaction between the two defects is exactly analogous to the Aharonov-Bohm interaction between a localized magnetic flux $\Phi$ and a localized electric charge $Q$ in two-spatial dimensions.  When a charge is adiabatically carried around a flux, the wave function of the system is modified by a phase $\exp(iQ\Phi/\hbar c)$ that is independent of the separation between charge and flux. Similarly, if a site defect is transported around a plaquette defect, the wave function of the system is modified by the phase $-1$ independent of the separation between the defects. Formally, this phase arises because of the anticommutation relation satisfied by $X$ and $Z$. Physically, it arises because the ground state of the system is very highly entangled and thus is able to support very long range quantum correlations. The protected qubits are encoded in the Aharonov-Bohm phases acquired by quasiparticles that travel around the fundamental nontrivial cycles of the surface; these could be measured in principle in a suitable quantum interference experiment.

It is useful to observe that the degeneracy of the ground state of the system is a necessary consequence of the unusual interactions among the quasiparticles \cite{einarsson,wen}. A unitary operator $U_{S,1}$ can be constructed that describes a process in which a pair of site defects is created, one member of the pair propagates around a nontrivial cycle $C_1$ of the surface, and then the pair reannihilates. Similarly a unitary operator $U_{P,2}$ can be constructed associated with a plaquette defect that propagates around a complementary nontrivial cycle $C_2$ that intersects $C_1$ once. These operators commute with the Hamiltonian $H$ of the system and can be simultaneously diagonalized with $H$, but $U_{S,1}$ and $U_{P,2}$ do not commute with one another. Rather, they satisfy (in an infinite system)
\begin{equation}
   {U_{P,2}}^{-1} ~ {U_{S,1}}^{-1}~U_{P,2}~U_{S,1}= -1 ~.
\end{equation}
The nontrivial commutator arises because the process in which (1) a site defect winds around $C_1$, (2) a plaquette defect winds around $C_2$ (3) the site defect winds around $C_1$ in the reverse direction, and (4) the plaquette defect winds around $C_2$ in the reverse direction, is topologically equivalent to a process in which the site defect winds once around the plaquette defect.

Because $U_{S,1}$ and $U_{P,2}$ do not commute, they cannot be simultaneously diagonalized --- indeed applying $U_{P,2}$ to an eigenstate of $U_{S,1}$ flips the sign of the $U_{S,1}$ eigenvalue. Physically, there are two distinct ground states that can be distinguished by the Aharonov-Bohm phase that is acquired when a site defect is carried around $C_1$; we can change this phase by carrying a plaquette defect around $C_2$.  Similarly, the operator $U_{S,2}$ commutes with $U_{S,1}$ and $U_{P,2}$ but anticommutes with $U_{P,1}$. Therefore there are four distinct ground states, labeled by their $U_{S,1}$ and $U_{S,2}$ eigenvalues.

This reasoning shows that the topological interaction between site defects and plaquette defects implies that the system on an (infinite) torus has a generic four-fold ground-state degeneracy. The argument is easily extended to show that the generic degeneracy on a genus $g$ Riemann surface is $2^{2g}$. By a further extension, we see that the generic degeneracy is $q^{2g}$ if the Aharonov-Bohm phase associated with winding one defect around another is 
\begin{equation}
\exp (2\pi ip/q)~,
\end{equation}
where $p$ and $q$ are integers with no common factor.

The same sort of argument can be applied to planar systems with a mass gap in which single defects can disappear at an edge. For example, consider an annulus in which site defects can disappear at the inner and outer edges. Then states can be classified by the Aharonov-Bohm phase acquired by a plaquette defect that propagates around the annulus, a phase that flips in sign if a site defect propagates from inner edge to outer edge. Hence there is a two-fold degeneracy on the annulus. For a disc with $h$ holes, the degeneracy is $2^h$ if site defects can disappear at any boundary, or $q^h$ if the Aharonov-Bohm phase of site defect winding about plaquette defect is $\exp(2\pi i p/q)$. 

These degeneracies are exact for the unperturbed model system, but will be lifted slightly in a weakly perturbed system of finite size. Loosely speaking, the effect of perturbations will be to give the defects a finite effective mass, and the lifting of the degeneracy is associated with quantum tunneling processes in which a virtual defect winds around a cycle of the surface. The amplitude $A$ for this process has the form 
\begin{equation}
A\sim C\exp\left(-\sqrt{2} (m^*\Delta)^{1/2}L/\hbar\right)~,
\end{equation}
where $L$ is the physical size of the shortest nontrivial (relative) cycle of the surface, $m^*$ is the defect effective mass, and $\Delta$ is the minimal energy cost of creating a defect.  The energy splitting is proportional to $A$, and like $A$ becomes negligible when the system is large compared to the characteristic length $l\equiv \hbar (m^*\Delta)^{-1/2}$. 

In this limit, and at sufficiently low temperature, the degenerate ground state provides a reliable quantum memory. 
If a pair of defects is produced by a thermal fluctuation, and one of the defects wanders around a nontrivial cycle before the pair reannihilates, then the encoded quantum information will be damaged. These fluctuations are suppressed by the Boltzman factor $\exp(-\Delta/kT)$ at low temperature. Even if defect nucleation occurs at a nonnegligible rate, we could enhance the performance of the quantum memory by continually monitoring the state of the defect gas. If the winding of defects around nontrivial cycles is detected and carefully recorded, damage to the encoded quantum information can be controlled.

\section{The statistical physics of error recovery}
\label{sec:statphys}

One of our main objectives in this paper is to invoke surface coding to establish an accuracy threshold for quantum computation --- how well must quantum hardware perform for quantum storage, or universal quantum computation, to be achievable with arbitrarily small probability of error? In this section, rather than study the efficacy of a particular fault-tolerant protocol for error recovery, we will address whether the syndrome of a surface code is adequate in principle for protecting quantum information from error. Specifically, we will formulate an order parameter that distinguishes two phases of a quantum memory: an ``ordered'' phase in which reliable storage is possible, and a ``disordered phase'' in which errors unavoidably afflict the encoded quantum information.  Of course, this phase boundary also provides an upper bound on the accuracy threshold that can be reached by any particular protocol. The toric code and the planar surface code have the same accuracy threshold, so we may study either to learn about the other.

\subsection{The error model}
\label{sec:error_model}

Let us imagine that in a single time step, we will execute a measurement of
each stabilizer operator at each site and each plaquette of the lattice.
During each time step, new qubit errors might occur.  To be concrete and to simplify the discussion, we assume that all qubit errors are stochastic, and so can be assigned probabilities. (For example, errors that arise from decoherence have this property.) We will also assume that the errors acting on different qubits are independent, that  bit-flip ($X$) errors and phase ($Z$) errors are uncorrelated with one another, and that $X$ and $Z$ errors are equally likely. Thus the error in each time step acting on a qubit with state $\rho$ can be represented by the quantum channel
\begin{eqnarray}
\rho \to & (1-p)^2 I\rho I + p(1-p) X\rho X  \nonumber\\
& +p(1-p) Z\rho X + p^2 Y\rho Y~,
\end{eqnarray}
where $p$ denotes the probability of either an $X$ error or a $Z$ error. It is easy to modify our analysis if some of these assumptions are relaxed; in particular, correlations between $X$ and $Z$ errors would not cause much trouble, since we have separate procedures for recovery from the $X$ errors and the $Z$ errors.

Faults can also occur in the syndrome measurement. We assume that these measurement errors are  uncorrelated. We will denote by $q$ the probability that the measured syndrome bit is faulty at a given site or plaquette.  

Aside from being uncorrelated in space, the qubit and measurement errors are
also assumed to be uncorrelated in time.  Furthermore, the qubit and
measurement errors are not correlated with one another.  We assume that $p$ and
$q$ are known quantities --- our choice of recovery algorithm depends on their
values. In Sec.~\ref{sec:storage_gates}, we will discuss how $p$ and $q$ can be related to more fundamental quantities, namely the fidelities of elementary quantum gates.  There we will see that the execution of the syndrome measurement circuit can introduce correlations between errors. Fortunately, these correlations (which we ignore for now) do not have a big impact on the accuracy threshold.

\subsection{Defects in spacetime}

Because syndrome measurement may be faulty, it is necessary to repeat the
measurement to improve our confidence in the outcome.  But since new errors may
arise during the repeated measurements, it is a subtle matter to formulate an
effective procedure for rejecting measurement errors.  

Let us suppose, for a toric block of arbitrarily large size, that we measure the error syndrome once per time step, that we monitor the block for an arbitrarily long time, and that we store all of the syndrome information that is collected. We want to address whether this syndrome information enables us to recover from errors with a probability of failure that becomes  exponentially small as the size of the toric block increases.  The plaquette check operators identify bit flips and the site check operators identify phase errors; therefore we consider bit-flip and phase error recovery separately. 

For analyzing how the syndrome information can be used most effectively, it is quite convenient to envision a {\em three-dimensional} simple cubic lattice, with the third dimension representing an integer-valued {\em time}.  We imagine that the error operation acts at each integer-valued time $t$, with a syndrome measurement taking place in between each $t$ and $t+1$. Qubits in the code block can now be associated with timelike
plaquettes, those lying in the $tx$ and $ty$ planes.  A qubit error that occurs
at time $t$ is associated with a horizontal (spacelike) link that lies in the
time slice labeled by $t$.  The outcome of the measurement of the stabilizer operator $X_s=X^{\otimes 4}=\pm1$ at site $s$, performed between time $t$ and time $t+1$, is marked on the vertical (timelike) link connecting site $s$ at time $t$ and site $s$ at time $t+1$. A similar picture applies to the history of the $Z_P$ stabilizer operators at each plaquette, but with the lattice replaced by its dual.

On some of these vertical links, the measured syndrome is erroneous.  We will
repeat the syndrome measurement $T$ times in succession, and the ``error
history'' can be described as a set of marked links on a lattice with altogether $T$ time slices. The error
history encompasses both error events that damage the qubits in the code block,
and faults in the syndrome measurements. On the initial ($t=0$) slice are
marked all uncorrected qubit errors that are left over from previous rounds of
error correction; new qubit errors that arise at a later time $t$
($t=1,2,\dots,T-1$) are marked on horizontal links on slice $t$. Errors in the
syndrome measurement that takes place between time $t$ and $t+1$ are marked on the corresponding vertical links. Errors on horizontal links occur with probability $p$, and
errors on vertical links occur with probability $q$.

\begin{figure}
\begin{center}
\leavevmode
\epsfxsize=3in
\epsfbox{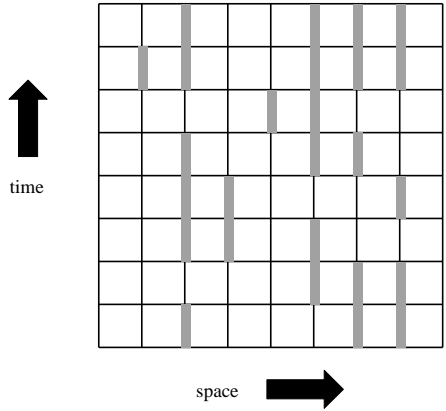}
\end{center}
\caption{The two-dimensional lattice depicting a history of the error syndrome for the quantum repetition code, with time running upward. Each row represents the syndrome at a particular time. Qubits reside on plaquettes, and two-qubit check operators are measured at each vertical link. Links where the syndrome is nontrivial are shaded.}
\label{fig:syndrome_history}
\end{figure}

\begin{figure}
\begin{center}
\leavevmode
\epsfxsize=3in
\epsfbox{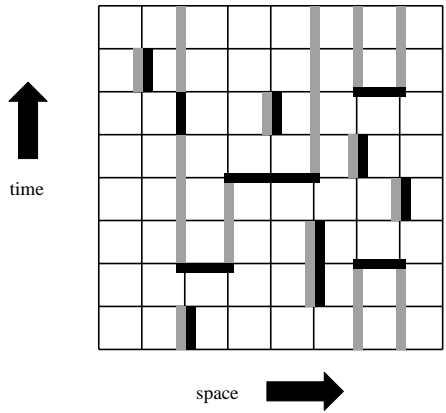}
\end{center}
\caption{An error history shown together with the syndrome history that it generates, for the quantum repetition code. Links where errors occured are darkly shaded, and links where the syndrome is nontrivial are lightly shaded. Errors on horizontal links indicate where a qubit flipped between successive syndrome measurements, and errors on vertical links indicate where the syndrome measurement was wrong. Vertical links that are shaded both lightly and darkly are locations where a nontrivial syndrome was found erroneously. The chain of lightly shaded links (the syndrome) and the chain of darkly shaded links (the errors) both have the same boundary}
\label{fig:error_history}
\end{figure}

For purposes of visualization, it is helpful to consider the simpler case of a quantum repetition code, which can be used to protect coherent quantum information from bit-flip errors if there are no phase errors (or phase errors if there are no bit-flip errors). In this case we may imagine that qubits reside on sites of a periodically identified one-dimensional lattice ({\it i.e.}, a circle); at each link the stabilizer generator $ZZ$ acts on the two neighboring sites. Then there is one encoded qubit --- the two-dimensional code space is spanned by the state $|000\dots 0\rangle$ with all spins ``up,'' and the state $|111\dots\rangle$ with all spins ``down.'' In the case where the syndrome measurement is repeated to improve reliability, we may represent the syndrome's history by associating qubits with plaquettes of a two-dimensional lattice, and syndrome bits with the timelike links, as shown in Fig.~\ref{fig:syndrome_history} and Fig.~\ref{fig:error_history}. Again, bit-flip errors occur on horizontal links with probability $p$ and syndrome measurement errors occur on vertical links with probability $q$. 

Of course, as already noted in Sec.~\ref{subsec:ftrec}, we may also use a two-dimensional lattice to represent the error configuration of the toric code, in the case where the syndrome measurements are perfect. In that case, we can collect reliable information by measuring the syndrome in one shot, and errors occur on links of the two-dimensional lattice with probability $p$.

\subsection{Error chains, world lines, and magnetic flux tubes}

In practice, we will always want to protect quantum information for some finite time. But for the purpose of investigating whether error correction will work effectively in principle, it is convenient to imagine that our repeated rounds of syndrome measurement extend indefinitely into the past and into the future. Qubit errors are continually occuring; as defects are created in pairs, propagate about on the lattice, and annihilate in pairs, the world lines of the defects form closed loops in spacetime. Some loops are homologically trivial and some are homologically nontrivial. Error recovery succeeds if we are able to correctly identify the homology class of each closed loop. But if a homologically nontrivial loop arises that we fail to detect, or if we mistakenly believe that a homologically nontrivial loop has been generated when none has been, then error recovery will fail. For now, let us consider this scenario in which we continue to measure the syndrome forever --- in Sec.~\ref{sec:finite_time}, we will consider some issues that arise when we perform error correction for a finite time. 

So let us imagine a particular history extending over an indefinite number of time slices, with the observed syndrome marked on each vertical link, measurement errors marking selected vertical links, and qubit errors marking selected horizontal links. For this history we may identify several distinct 1-chains (sets of links). We denote by $S$ the {\em syndrome chain} containing all (vertical) links at which the measured syndrome is nontrivial ($X_s=-1$). We denote by $E$ the {\em error chain} containing all links where errors have occurred, including both qubit errors on horizonal links and measurement errors on vertical links. Consider $S+E$, the disjoint union of $S$ and $E$ ($S+E$ contains the links that are in either $S$ or $E$, but not both). The chain $S+E$  represents the ``actual'' world lines of the defects generated by qubit errors, as illustrated in Fig.~\ref{fig:error_history}. Its vertical links are those on which the syndrome would be nontrivial were it measured without error. Its horizontal links are events where a defect pair is created, a pair annihilates, or an existing defect propagates from one site to a neighboring site. Since the world lines never end, the chain $S+E$ has no boundary, $\partial(S+E)=0$. Equivalently $S$ and $E$ have the same boundary, $\partial S=\partial E$. 

Hence, the measured syndrome $S$ reveals the boundary of the error chain $E$; we may write $E=S+C$, where $C$ is a {\em cycle} (a chain with no boundary). But any other error chain $E'=S+C'$, where $C'$ is a cycle, has the same boundary as $E$ and therefore could have caused the same syndrome. To recover from error, we will use the syndrome information to make a hypothesis, guessing that the actual error chain was $E'=S+C'$. Now, $E'$ may not be the same chain as $E$, but as long as the cycle $E+E'=C+C'$ is homologically trivial (the boundary of a surface) then recovery will be successful. If $C+C'$ is homologically nontrivial, then recovery will fail. We say that $C$ and $C'$ are in the same {\em homology class} if $C+C'$ is homologically trivial. Therefore, whether we can protect against error hinges on our ability to identify, not the cycle $C$, but rather the homology class of $C$.

Considering the set of all possible histories, let ${\rm prob}(E')$ denote the probability of the error chain $E'$ (strictly speaking, we should consider the total elapsed time to be finite for this probability to be defined). Then the probability that the syndrome $S$ was caused by any error chain $E'=S+C'$, such that $C'$ belongs to the homology class $h$, is
\begin{equation}
{\rm prob}(h|S)={\sum_{C'\in h}{\rm prob}(S+C')\over \sum_{C'}{\rm prob}(S+C')}
\label{prob_homol}
\end{equation}
Clearly, then, given a measured syndrome $S$, the optimal way to recover is to guess that the homology class $h$ of $C$ is the class with the highest probability according to eq.~(\ref{prob_homol}). Recovery succeeds if $C$ belongs to this class, and fails otherwise.

We say that the probability of error per qubit lies below the accuracy threshold if and only if 
the recovery procedure fails with a probability that vanishes as the linear size $L$ of the lattice increases to infinity. Therefore, below threshold, the cycle $C$ actually belongs to the class $h$ that maximizes eq.~(\ref{prob_homol}) with a probability that approaches one as $L\to\infty$. It is convenient to restate this criterion in a different way that makes no explicit reference to the syndrome chain $S$. We may write the relation between the actual error chain $E$ and the hypothetical error chain $E'$ as $E'=E+D$, where $D$ is the cycle that we called $C+C'$ above. Let ${\rm prob}[(E+D)|E]$ denote the normalized conditional probability for error chains $E'=E+D$ that have the same boundary as $E$.  Then, the probability of error per qubit lies below threshold if and only if, in the limit $L\to \infty$,
\begin{equation}
\sum_E {\rm prob}(E)\cdot \sum_{D~{\rm nontrivial}} {\rm prob}[(E+D)|E]=0~.
\label{ec_cond}
\end{equation}
Eq.~(\ref{ec_cond}) says that error chains that differ from the actual error chain by a homologically nontrivial cycle have probability zero. Therefore, the observed syndrome $S$ is sure to point to the correct homology class, in the limit of an arbitrarily large code block.

This accuracy threshold achievable with toric codes can be identified with a phase transition in a particular statistical-physics model defined on a lattice. In a sense that we will make precise, the error chains are analogous to magnetic flux tubes in a superconductor, and the boundary points of the error chains are magnetic monopoles where these flux tubes terminate. Fixing the syndrome pins down the monopoles, and the ensemble of chains with a specified boundary can be regarded as a thermal ensemble. As the error probability increases, the thermal fluctuations of the flux tubes increase, and at the critical temperature corresponding to the accuracy threshold, the flux tubes condense and the superconductivity is destroyed. 

A similar analogy applies to the case where the syndrome is measured perfectly, and a two-dimensional system describes the syndrome on a single time slice. Then the error chains are analogous to domain walls in an Ising ferromagnet, and the boundary points of the error chains are ``Ising vortices'' where domain walls terminate.  Fixing the syndrome pins down the vortices, and the ensemble of chains with a specified boundary can be interpreted as a thermal ensemble. As the error probability increases, the domain walls heat up and fluctuate more vigorously. At a critical temperature corresponding to the accuracy threshold, the domain walls condense and the system becomes magnetically disordered. This two-dimensional model also characterizes the accuracy threshold achievable with a quantum repetition code, if the syndrome is imperfect and the qubits are subjected only to bit-flip errors (or only to phase errors). 

\subsection{Derivation of the model}

Let us establish the precise connection between our error model and the corresponding
statistical-physics model . In the two-dimensional case, we consider a
square lattice with links representing
qubits, and assume that errors arise independently on each link with probability $p$.
In the three-dimensional case, we consider a simple cubic lattice. Qubits reside on the timelike plaquettes, and qubit errors arise independently with probability $p$ on spacelike links. Measurement errors occur independently with probability $q$ on timelike links. For now, we will make the simplifying assumption that $q=p$ so that the model is isotropic; the generalization to $q\ne p$ is straightforward.

An error chain $E$, in either two or three dimensions, can be characterized by a function
$n_E(\ell) $ that takes a link $\ell$ to $n_E(\ell)\in \{0, 1\}$, where
$n_E(\ell) = 1$ for each link $\ell$ that is occupied by the chain.  Hence
the probability that error chain $E$ occurs is
\begin{eqnarray}
{\rm prob}(E) &=& {\prod_\ell} (1 - p)^{1 - n_E(\ell)} p^{n_E(\ell)}
\nonumber \\
&=& \left[{\prod_\ell} (1 - p)\right] \cdot {\prod_\ell} \left(\frac{p}{1 -
p}\right)^{n_E(\ell)}~,
\end{eqnarray}
where the product is over all links of the lattice.

Now suppose that the error chain $E$ is fixed, and we are interested in
the probability distribution for all chains $E'$ that have the same
boundary as $E$.  Note that we may express $E' = E + C$, where
$C$ is a cycle (a chain with no boundary) and consider the probability
distribution for $C$.  Then if $n_C(\ell) = 1$ and $n_E(\ell) = 0$,
the link $\ell$ is occupied by $E'$ but not by $E$, an event whose
probability (aside from an overall normalization) is
\begin{equation}
\left(\frac{p}{1-p}\right)^{n_C(\ell)}~.
\end{equation}
But if $n_C(\ell) = 1$ and $n_E(\ell) = 1$, then the link $\ell$ is not
occupied by $E'$, an event whose probability (aside from an overall normalization) is
\begin{equation}
\left(\frac{1-p}{p}\right)^{n_C(\ell)}.
\end{equation}
Thus a chain $E'=E + C$ with the same
boundary as $E$ occurs with probability 
\begin{equation}
{\rm prob}(E'|E) \propto {\prod_\ell} \exp\left(J_\ell u_\ell\right)~;
\end{equation}
here we have defined
\begin{equation}
u_\ell= 1 - 2 n_C(\ell) \in \{1, -1\},
\end{equation}
and the coupling $J_\ell$ assigned to link $\ell$ has the form
\begin{equation}
e^{-2J_{\ell}} = \cases{
p/(1-p), & {for} $\ell \not\in E$,\cr
(1-p)/p, &  {for} $\ell \in E$.\cr}
\end{equation}
Recall that the 1-chain $\{\ell| u_\ell=-1\}$ is required to be a {\em
cycle} --- it has no boundary.

It is obvious from this construction that ${\rm prob}(E'|E)$ does not depend on
how the chain $E$ is chosen --- it depends only on the boundary of $E$.
We will verify this explicitly below.

The cycle condition satisfied by the $u_\ell$'s can be expressed as
\begin{equation}
\prod_{\ell \ni s} u_\ell = 1~;
\end{equation}
at each site $s$, an even number of links incident on that site have $u_\ell = -
1$.  It is convenient to {\em solve} this condition, expressing
the $u_\ell$'s in terms of unconstrained variables.  To achieve this in two dimensions, we associate with each link $\ell$ a link $\ell^*$ of the {\em dual lattice.}  Under
this duality, sites are mapped to plaquettes, and the cycle condition
becomes
\begin{equation}
\prod_{\ell^*\in P^*} u_{\ell^{*}}= 1.
\end{equation}
To solve the constraint, we introduce variables $\sigma_i \in \{1,-1\}$ associated with each
site $i$ of the dual  lattice, and write
\begin{equation}
u_{ij} = \sigma_i \sigma_j
\end{equation}
where $i$ and $j$ are nearest-neighbor sites.

Our solution to the constraint is not quite the most general possible.
In the language of differential forms, we have solved the condition
$du = 0$ (where $u$ is a discrete version of a one-form, and $d$ denotes the exterior derivative) by
writing $u = d \sigma$, where $\sigma$ is a zero-form.  Thus our solution
misses the cohomologically nontrivial closed forms, those that are not
exact.  In the language of homology, our solution includes all and only
those cycles that are homologically trivial --- that is, cycles that bound  a
surface.

In three dimensions, links are dual to plaquettes, and sites to cubes.  The cycle
condition becomes, on the dual lattice,
\begin{equation}
\prod_{P^* \in C^*} u_{P^{*}} = 1~;
\end{equation}
each dual cube $C^*$ contains an even number of dual plaquettes that are
occupied by the cycle.  We solve this constraint by introducing
variables $\sigma_{\ell^{*}} \in\{1,-1\}$ on the dual links, and
defining
\begin{equation}
u_{P^{*}} = \prod_{\ell^* \in P^*} \sigma_{\ell^{*}}.
\end{equation}
In this case, we have solved a discrete version of $du = 0$, where $u$
is a two-form, by writing $u = d\sigma$, where $\sigma$ is a one-form.
Once again, our solution generates only the cycles that are
homologically trivial.

We have now found that, in two dimensions, the ``fluctuations'' of the error chains
$E'$ that share a boundary with the chain $E$ are described by a
statistical-mechanical model with partition function
\begin{equation}
Z[J,\eta] = \sum_{\{\sigma_i\}} \exp \left(J {\sum_{\langle
ij\rangle}} ~\eta_{ij} \sigma_i \sigma_j\right),
\end{equation}
where $e^{-2J} = p/(1-p)$. The sum in the exponential is over pairs of nearest neighbors on a
square lattice, and $\eta_\ell \in \{1, -1\}$ is defined by
\begin{equation}
\eta_\ell = \cases{
1, & if $\ell\not\in E^*$,\cr
-1 & if $\ell \in E^*$.\cr}
\end{equation}
Furthermore if the error chains $E$ and $E'$ are generated by sampling the
same probability distribution, then the $\eta_\ell$'s are chosen at
random subject to
\begin{equation}
\eta_\ell = \cases{
1, & with probability $1-p$,\cr
-1 & with probability $p$.\cr}
\end{equation}
This model is the well-known ``random-bond Ising model.''  Furthermore,
the relation $e^{-2J} = p/(1-p)$ between the coupling and the bond
probability defines the ``Nishimori line'' \cite{nish_line} in the phase diagram of the
model, which has attracted substantial attention\footnote{For a recent discussion, see \cite{gruzberg}.} because the model is
known to have enhanced symmetry properties on this line.

\begin{figure}
\begin{center}
\leavevmode
\epsfxsize=3in
\epsfbox{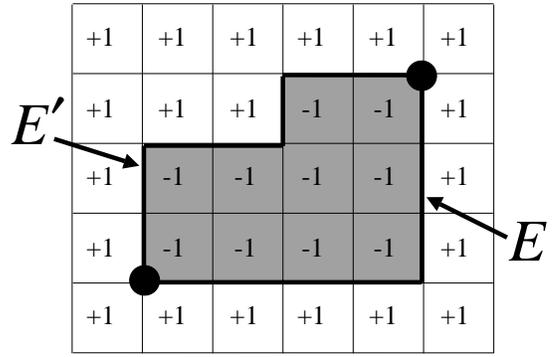}
\end{center}
\caption{The ``quenched'' error chain $E$ and the ``fluctuating'' error chain $E'$, as represented in the two-dimensional random-bond Ising model. Ising spins taking values in $\{\pm 1\}$ reside on plaquettes, Ising vortices are located on the sites marked by filled circles, and the coupling between neighboring spins is antiferromagnetic along the path $E$ that connects the Ising vortices. The links of $E'$ comprise a domain wall connecting the vortices. The closed path $C=E+E'$ encloses a domain of spins with the value $-1$. }
\label{fig:ising_error}
\end{figure}

Perhaps the interpretation of this random-bond Ising model can be grasped better if we picture the original lattice rather than the dual lattice, so that the Ising spins reside on plaquettes as in Fig.~\ref{fig:ising_error}. The coupling between spins on neighboring plaquettes is antiferromagnetic on the links belonging to the chain $E$ (where $\eta_\ell=-1$), meaning that it is energetically preferred for the spins to antialign at these links. At links not in $E$ (where $\eta=1$), it is energetically preferred for the spins to align. Thus a link $ij$ is excited if $\eta_{ij} \sigma_i \sigma_j = - 1$.  We say that the excited links constitute ``domain walls.''  In the case where $\eta_\ell =1$ on every link, a wall marks the boundary between two regions in which the spins point in opposite directions.  Walls can never end, because the boundary of a boundary is zero.

But if the $\eta$ configuration is nontrivial then the ``walls'' can
end.  Indeed each boundary point of the chain $E$ of links with $\eta_\ell =
- 1$ is an endpoint of a wall, what we will call an ``Ising vortex.''  For example, for the configuration shown in Fig~\ref{fig:ising_error}, a domain wall occupies the chain $E'$ that terminates on Ising vortices at the marked sites. The figure also illustrates that the model depends only on the boundary of the chain $E$, and not on other properties of the chain.  To see this, imagine performing the change of variables
\begin{equation}
\sigma_i \rightarrow -\sigma_i,
\end{equation}
on the shaded plaquettes of Fig.~\ref{fig:ising_error}. A mere change of variable cannot alter the locations of the excited links --- rather the effect is to shift the antiferromagnetic couplings from the chain $E$ to a different chain $E'$ with the same boundary.

In three dimensions, the fluctuations of the error chains that share a boundary with
the specified chain $E$ are described by a model with partition function
\begin{equation}
Z[J,\eta] = \sum_{\{\sigma_\ell\}} \exp (J \sum_P \eta_P u_P),
\end{equation}
where $u_P = \prod_{\ell\in P} \sigma_\ell$ and
\begin{equation}
\eta_P = \cases{
1, & if $P\not\in E^*$,\cr
-1, & if $ P\in E^*$.\cr}
\end{equation}
This model is a ``random-plaquette'' $Z_2$ gauge theory in three
dimensions, which, as far as we know, has not been much studied
previously.  Again, we are interested in the ``Nishimori line'' of this
model where $e^{-2J} = p/(1-p)$, and $p$ is the probability that a
plaquette has $\eta_P = - 1$.

In this three-dimensional model, we say that a pla\-quette $P$ is excited if $\eta_P u_P =
- 1$.  The excited pla\-quettes constitute ``magnetic flux
tubes'' --- these form closed loops on the original lattice if $\eta_P = 1$ on
every plaquette.  But at each boundary point of the chain $E$ on the original lattice (each cube on the dual lattice that contains an odd number of plaquettes with $\eta_{P} = - 1$),  the flux tubes can end.  The sites of the original  lattice (or cubes of the dual lattice) that
contain endpoints of magnetic flux tubes are said to be ``magnetic monopoles.''

\subsection{Order Parameters}

As noted, our statistical-mechanical model includes a sum over those and
only those chains $E'$ that are {\em homologically equivalent} to the chain
$E$.  To determine whether errors can be corrected reliably, we want to
know whether chains $E'$ in a {\em different} homology class than $E$ have
negligible probability in the limit of a large lattice (or code block).
The relative likelihood of different homology classes is determined by the free energy
difference of the classes; in the ordered phase, we anticipate that the
free energy of nontrivial classes exceeds that of the trivial classes by
an amount that increases linearly with $L$, the linear size of the
lattice.

But for the purpose of finding the value of the error probability at the
accuracy threshold, it suffices to consider the model in an infinite
volume (where there is no nontrivial homology).  In the ordered phase
where errors are correctable, large fluctuations of domain walls or flux
tubes are suppressed, while in the disordered phase the walls or tubes
``dissolve'' and cease to be well defined.

Thus, the phase transition corresponding to the accuracy threshold is a singularity, in the infinite-volume limit, in the ``quenched'' free energy, defined as
\begin{equation}
\langle \beta F[J,\eta] \rangle_p\equiv -\sum_{\{\eta\}}{\rm Prob}(\eta)\cdot \ln Z[J,\eta]~,
\end{equation}
where
\begin{equation}
{\rm Prob}(\eta)= \prod_\ell (1-p)^{1-\eta_\ell}p^{\eta_\ell}
\end{equation}
in two dimensions, or 
\begin{equation}
{\rm Prob}(\eta)= \prod_P (1-p)^{1-\eta_P}p^{\eta_P}
\label{eta_prob}
\end{equation}
in three dimensions. The term ``quenched'' signifies that, although the $\eta$ chains are generated at random, we consider thermal fluctuations with the positions of the vortices or monopoles pinned down. The inverse temperature $\beta$ is identical to the coupling $J$. We use the notation $\langle\cdot\rangle_p$ to indicate an average with respect to the quenched randomness, and we will denote by $\langle \cdot\rangle_\beta$ an average over thermal fluctuations.

There are various ways to describe the phase transition
in this system, and to specify an order parameter. For example, in the two-dimensional Ising system, we may consider a ``disorder parameter'' $\Phi(x)$ that inserts
a single Ising vortex at a specified position $x$.  To define this
operator, we must consider either an infinite system or a finite system
with a boundary; on the torus, Ising vortices can only be inserted in
pairs.  But for a system with a boundary, we can consider a domain wall
with one end at the boundary and one end in the bulk.  In the {\em ferromagnetic} phase, the cost in free energy of introducing an
additional vortex at $x$ is proportional to $L$, the distance from $x$ to the
boundary.  Correspondingly we find
\begin{equation}
\langle\langle \Phi (x)\rangle_\beta \rangle_p= 0
\end{equation}
in the limit $L \rightarrow \infty$.  The disorder parameter vanishes because we cannot
introduce an isolated vortex without creating an infinitely long domain
wall.  In the disordered phase, an additional vortex can be introduced
at finite free energy cost, and hence
\begin{equation}
\langle\langle \Phi (x)\rangle_\beta\rangle_p \not= 0~.
\end{equation}

On the torus, we may consider an operator that inserts, not a
semi-infinite domain wall terminating on a vortex, but instead a domain
wall that winds about a cycle of the torus.  Again, in the
ferromagnetically ordered phase, the cost in free energy of inserting
the domain wall will be proportional to $L$, the minimal length of a
cycle.  Specifically, in our two-dimensional
Ising spin model, consider choosing an $\eta$-chain and
evaluating the corresponding partition function
\begin{equation}
Z[J,\eta] = \exp [-\beta F(J,\eta)]~.
\end{equation}
Now choose a set of links $C$ of the original lattice that constitute a
nontrivial cycle wound around the torus, and replace $\eta_\ell
\rightarrow -\eta_\ell$ for the corresponding links of the dual lattice,
$\ell \in C^*$.  Evaluate, again, the partition function, obtaining
\begin{equation}
Z_C [J,\eta] = \exp [-\beta F_C (J,\eta)]~.
\end{equation}
Then the free energy cost of the domain wall is given by
\begin{equation}
\beta F_C (J,\eta) - \beta F(J,\eta) = - \ln \left(\frac{Z_C [J,\eta]}{Z [J,\eta]}\right)~.
\end{equation}
After averaging over $\{\eta\}$, this free energy cost diverges as $L\to\infty$ in the ordered phase, and converges to a constant in the disordered phase.

There is also a dual order parameter that vanishes in the disordered 
phase --- the spontaneous magnetization of the Ising spin system. Strictly speaking, the defining property of the non-ferromagnetic disordered phase is that spin correlations decay with distance, so that
\begin{equation}
\lim_{r\to \infty}\langle\langle \sigma_0\sigma_r\rangle_\beta\rangle_p=0
\end{equation}
in the disordered phase. Correspondingly, the mean squared magnetization per site
\begin{equation}
m^2\equiv N^{-2}\sum_{i,j}\langle\langle \sigma_i\sigma_j\rangle_\beta\rangle_p~,
\end{equation}
where $i,j$ are summed over all spins and $N$ is the total number of spins, approaches a nonzero constant as $N\to \infty$ in the ordered phase, and approaches zero as a positive power of $1/N$ in the disordered phase.

Similarly in our three-dimensional gauge theory, there is a disorder parameter that
inserts a single magnetic monopole, which we may think of as the end of a
semi-infinite flux tube.  Alternatively, we may consider the free energy cost of inserting a flux tube that wraps around the torus, which is proportional to $L$ in the magnetically ordered phase. In the three-dimensional model, the partition function $Z_C [J,\eta]$ in the presence of a flux tube wrapped around the nontrivial cycle $C$ of the original lattice is obtained by replacing $\eta_P \rightarrow - \eta_P$ on the plaquettes dual to
the links of $C$. The magnetically ordered phase is called a ``Higgs
phase'' or a ``superconducting phase.''  The magnetically disordered
phase is called a ``confinement phase'' because in this phase
introducing an isolated electric charge has a infinite cost in free energy, and
electric charges are confined in pairs by electric flux tubes. 

An order parameter for the Higgs-confinement transition is the Wilson loop operator
\begin{equation}
W(C)=\prod_{\ell\in C} \sigma_\ell
\end{equation}
associated with a closed loop $C$ of links on the lattice. This operator can be interpreted as the insertion of a charged particle source whose world line follows the path $C$. In the confinement phase, this world line becomes the boundary of the world sheet of an electric flux tube, so that the free energy cost of inserting the source is proportional to the minimal area of a surface bounded by $C$; that is, 
\begin{equation}
-\langle\ln \langle W(C)\rangle_\beta\rangle_p
\end{equation}
increases like the area enclosed by the loop $C$ in the confinement phase, while in the Higgs phase it increases like the perimeter of $C$.\footnote{A subtle point is that the relevant Wilson loop operator differs from that considered in Sec. 10 of \cite{alford}. In that reference, the Wilson loop was modified so that the ``Dirac strings'' connecting the monopoles would be invisible. But in our case, the Dirac strings have a physical meaning (they comprise the chain $E$) and we are genuinely interested in how far the physical flux tubes (comprising the chain $E'$) fluctuate away from the Dirac strings!}

In the case $q\ne p$, our gauge theory becomes anisotropic --- $p$ controls the coupling and the quenched disorder on the timelike plaquettes, while $q$ controls the coupling and the quenched disorder on the spacelike plaquettes. The tubes of flux in $E+E'$ will be stretched in the time direction for $q > p$ and compressed in the time direction for $q < p$. Correspondingly, spacelike and timelike Wilson loops will decay at different rates. Still, one expects that (for $0<q<1/2$) a single phase boundary in the $p$--$q$ plane separates the region in which both timelike and spacelike Wilson loops decay exponentially with  area (confinement phase) from the region in which both timelike and spacelike Wilson loops decay exponentially with perimeter. In the limit $q\to 0$, flux on the spacelike plaquettes becomes completely suppressed, and the timelike plaquettes on distinct time slices decouple, each described by the two-dimensional spin model described earlier. Similarly, in the limit $p\to 0$, the gauge theory reduces to decoupled one-dimensional spin models extending in the vertical direction, with a critical point at $q=1/2$.

\subsection {Accuracy threshold}
\label{subsec:acc_thresh}

What accuracy threshold can be achieved by surface codes? We have found that in the case where the syndrome is measured perfectly ($q=0$), the answer is determined by the value of critical point of the two-dimensional random-bond Ising model on the Nishimori line. This value has been determined by numerically evaluating the domain wall free energy; a recent result of Honecker et al. is \cite{honecker}
\begin{equation}
\label{eq:honecker}
p_c = .1094\pm .0002~.
\end{equation}

A surface code is a Calderbank-Shor-Steane (CSS) code, meaning that each stabilizer generator is either a tensor product of $X$'s or a tensor product of $Z$'s \cite{cal_shor,steane_css}. If $X$ errors and $Z$ errors each occur with probability $p$, then it is known that CSS codes exist with asymptotic rate $R\equiv k/n$ (where $n$ is the block size and $k$ is the number of encoded qubits) such that error recovery will succeed with probability arbitrarily close to one, where  
\begin{equation}
R = 1 - 2H_2(p)~;\end{equation}
here $H_2(p)= -p\log_2 p -(1-p)\log_2(1-p)$ is the binary Shannon entropy. This rate hits zero when $p$ has the value
\begin{equation}
p_c = .1100 ~,
\end{equation}
which agrees with eq.~(\ref{eq:honecker}) within statistical errors. Thus the critical error probability is (at least approximately) the same regardless of whether we allow arbitrary CSS codes or restrict to those with a locally measurable syndrome. This result is analogous to the property that the classical repetition code achieves reliable recovery from bit-flip errors for any error probability $p<1/2$, the value for which the Shannon capacity hits zero.
Note that eq.~(\ref{eq:honecker}) can also be interpreted as a threshold for the quantum repetition code, in the case where the bit-flip error rate and the measurement error rate are equal ($p=q$).

If measurement errors are incorporated, then the accuracy threshold achievable with surface codes is determined by the critical point along the Nishimori line of the three-dimensional $Z_2$ gauge theory with quenched randomness. In that model the
measurement error probability $q$ (the error weight for vertical links)
and the bit-flip probability $p$ (the error weight for horizontal links)
are independent parameters.  It seems that numerical studies of this quenched gauge theory have not been done previously, even in the isotropic case; work on this problem is in progress. 

Since recovery is more difficult with imperfect syndrome information than with perfect syndrome information, the numerical data on the random-bond Ising model indicate that $p_c < .11$ for any $q>0$. For the case $p=q$, we will derive the lower bound $p_c\ge .0114$ in Sec.~\ref{sec:chains_minimal}.

\subsection{Free energy versus energy}
\label{free_energy_energy}

In either the two-dimensional model (if $q=0$) or the three-dimensional model (if $q>0$), the critical error probability along the Nishimori line provides a criterion for whether it is possible in principle to perform
flawless recovery from errors.  In practice, we would have to execute a
classical computation, with the measured syndrome as input, to determine
how error recovery should proceed.  The defects revealed by the syndrome
measurement can be brought together to annihilate in several
homologically distinct ways; the classical computation determines which
of these ``recovery chains'' should be chosen.  

We can determine the right homology class by computing the free energy
for each homology class, and choosing the one with minimal free energy.
In the ordered phase (error probability below threshold) the correct sector 
will be separated in free energy from other sectors by an amount linear in $L$, the linear size of the lattice.

\begin{figure}
\begin{center}
\leavevmode
\epsfxsize=3.25in
\epsfbox{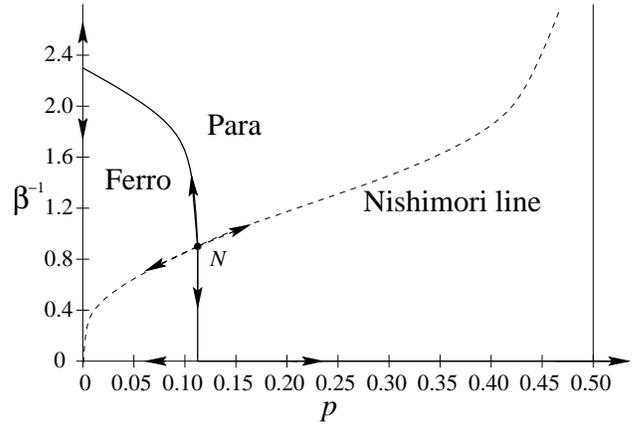}
\end{center}
\caption{The phase diagram of the random-bond Ising model, with the temperature $\beta^{-1}$ on the vertical axis and the probability $p$ of an antiferromagnetic bond on the horizontal axis. The solid line is the boundary between the ferromagnetic (ordered) phase and the paramagnetic (disordered) phase. The dotted line is the Nishimori line $e^{-2\beta}= p/(1-p)$, which crosses the phase boundary at the Nishimori point $N$. From the point $N$ to the horizontal axis, the phase boundary is {\em vertical}.}
\label{fig:nishi}
\end{figure}

The computation of the free energy could be performed by, for example, the Monte Carlo method. It should be possible to identify the homology class that minimizes the free energy in a time polynomial in $L$, unless the equilibration time of the
system is exponentially long.  Such a long equilibration time would be
associated with spin-glass behavior --- the existence of a large number
of metastable configurations.  In the random-bond Ising model, spin glass
behavior is expected in the disordered phase, but not in the
ferromagnetically ordered phase corresponding to error probability below
threshold.  Thus, we expect that in the two-dimensional model the
correct recovery procedure can be computed efficiently for any $p <
p_c$. Similarly, it is also reasonable to expect that, for error probability below threshold, the correct recovery chain can be found efficiently in the three-dimensional model that incorporates measurement errors.

In fact, there is reason to expect that when the error probability is below threshold, we can recover successfully by finding a recovery chain that minimizes {\em energy} rather than free energy. Nishimori \cite{nish_vert}  notes that along the Nishimori line, the free energy $\langle\beta F[J]\rangle_p$ coincides with the {\em entropy of frustration}; that is, the {\em Shannon entropy} of the distribution of Ising vortices. (He considered the isotropic two-dimensional model, but his argument applies just as well to our three-dimensional gauge theory, or to the anisotropic model with $q\ne p$.) Thus, the singularity of the free energy on the Nishimori line can be regarded as a singularity of this Shannon entropy, which is a purely geometrical effect having nothing to do with thermal fluctuations. 

On this basis, we may expect that there is a vertical phase boundary in our model, occurring at a fixed value of $p$ for all temperatures below the critical temperature at the Nishimori point, as indicated in Fig.~\ref{fig:nishi}; for the two-dimensional random-bond Ising model, this expectation has been reasonably well confirmed by numerical computations. Thus, the critical error probability can be computed by analyzing the phase transition at zero temperature, where the thermal entropy of the fluctuating chains can be neglected. In other words, in the ordered phase, the chain of minimal energy with the same boundary as the actual error chain will with probability one be in the same homology class as the error chain, in the infinite-volume limit. Ordinarily, minimizing free energy and energy are quite different procedures that give qualitatively distinct results. What seems to make this case different is that the quenched disorder (the error chain $E$) and the thermal fluctuations (the error chain $E'$) are drawn from the same probability distribution. 

Minimizing the energy has advantages. For one, the minimum energy configuration is the minimum weight chain with a specified boundary, which we know can be computed in a time polynomial in $L$ using the perfect matching algorithm of Edmonds \cite{edmonds,barahona_match}. Kawashima and Aoki \cite{kawashima} computed the energetic cost of introducing a domain wall at zero temperature, and found $p_c \simeq .105 \pm .002$, which is marginally consistent with the value  $p_c\simeq .1094\pm .0002$ computed by Honecker et al. at the Nishimori point \cite{honecker}.

Minimizing the energy is easier to analyze than minimizing the free energy, and at the very least the critical value of $p$ at zero temperature provides a {\em lower bound} on $p_c$ along the Nishimori line. In Sec.~\ref{sec:chains_minimal} we will derive a rigorous bound on the accuracy threshold in our error model, by considering the efficacy of the energy minimization procedure in the three-dimensional model.

\section{Chains of minimal weight}
\label{sec:chains_minimal}

\subsection{The most probable world line}

As argued in Sec.~\ref{free_energy_energy}, an effective way use the error syndrome in our three-dimensional model is to construct an error chain that has the minimal ``energy'' --- that is, we select from among all error chains that have the same boundary as the syndrome chain $S$, the single chain $E_{\rm min}$ that has the highest probability. In this Section, we will study the efficacy of this procedure, and so obtain a lower bound on the accuracy threshold for quantum storage.

An error chain $E$ with $H$ horizontal links and $V$ vertical links occurs with probability (aside from an overall normalization)
\begin{equation}
\left({p\over 1-p}\right)^{H}\left({q\over 1-q}\right)^{V}~,
\end{equation}
where $p$ is the qubit error probability and $q$ is the measurement error
probability.
Thus we choose $E_{\rm min}$ to be the chain with
\begin{equation}
\partial E_{\rm min}=\partial S
\end{equation}
that has the {\em minimal} value of
\begin{equation}
\label{eq:weight}
H\cdot \log \left({1-p\over p}\right) + V\cdot\log \left({1-q\over q}\right)~;
\end{equation}
we minimize the effective length (number of links) of the chain, but with
horizontal and vertical links given different linear weights for $p\ne q$.  If
the minimal chain is not unique, one of the minimal chains is selected
randomly.

Given the measured syndrome, and hence its boundary $\partial S$, the minimal chain $E_{\rm min}$ can be determined on a
classical computer, using standard algorithms, in a time bounded by a
polynomial of the number of lattice sites \cite{edmonds,barahona_match}.  If $p$ and $q$
are small, so that the lattice is sparsely populated by the sites contained in
$\partial S$, this algorithm typically runs quite
quickly. We assume this classical computation can be performed instantaneously and
flawlessly.

\subsection{A bound on chain probabilities}
Recovery succeeds if our hypothesis $E_{\rm min}$ is homologically equivalent to the actual error chain $E$ that generated the syndrome chain $S$, and fails otherwise. Hence, we wish to bound the likelihood of
homologically nontrivial paths appearing in $E+E_{\rm min}$.

Consider a particular cycle on our spacetime lattice (or in fact any connected path, whether or not the path is closed). Suppose
that this path contains $H$ horizontal links and $V$ vertical links.
How likely is it that $E+E_{\rm min}$ contains this particular set of links?

\begin{figure}
\begin{center}
\leavevmode
\epsfxsize=3in
\epsfbox{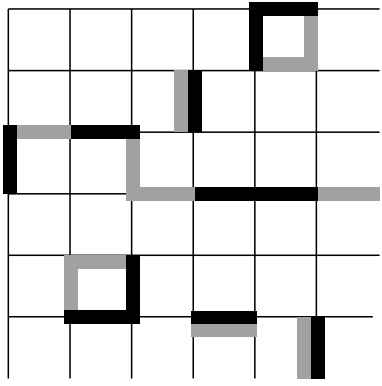}
\end{center}
\caption{The error chain $E$ (darkly shaded) and one possible choice for the chain $E_{\rm min}$ (lightly shaded), illustrated for a $6\times 6$ torus in two dimensions. In this case $E+E_{\rm min}$ contains a homologically nontrivial cycle of length 8, which contains $H_e=4$ links of $E$ and $H_m=4$ links of $E_{\rm min}$.}
\label{fig:min_chain}
\end{figure}

For our particular path with $H$ horizontal links and $V$ vertical links, let
$H_m,V_m$ be the number of those links contained in $E_{\rm min}$, and let
$H_e,V_e$ be the number of those links contained in $E$ (Cf. Fig.~\ref{fig:min_chain}). These quantities obey the relations
\begin{equation}
H_m+H_e\ge H~,\quad V_m + V_e \ge V~,
\end{equation}
and so it follows that
\begin{eqnarray}
\label{e_m_ineq1}
\left({p\over 1-p}\right)^{H_m}\left({q\over 1-q}\right)^{V_m} \cdot \left({p\over 1-p}\right)^{H_e}\left({q\over 1-q}\right)^{V_e} \nonumber\\
\le \left({p\over 1-p}\right)^{H}\left({q\over 1-q}\right)^{V}~.
\end{eqnarray}
Furthermore, our procedure for constructing $E_{\rm min}$ ensures that
\begin{equation}
\label{e_m_ineq2}
\left({p\over 1-p}\right)^{H_e}\left({q\over 1-q}\right)^{V_e}\le \left({p\over 1-p}\right)^{H_m}\left({q\over 1-q}\right)^{V_m}  ~.
\end{equation}
This must be so because the $e$ links and the $m$ links share the same 
boundary; were eq.~(\ref{e_m_ineq2}) not satisfied, we could replace the $m$
links in $E_{\rm min}$ by the $e$ links and thereby increase the value of $[p/(1-p)]^{H_{m}}[q/(1-q)]^{V_{m}}$.  Combining the inequalities eq.~(\ref{e_m_ineq1}) and
eq.~(\ref{e_m_ineq2}) we obtain
\begin{equation}
\label{e_ineq}
\left({p\over 1-p}\right)^{H_e}\left({q\over 1-q}\right)^{V_e}\le \left[\left({p\over 1-p}\right)^{H}\left({q\over 1-q}\right)^{V}\right]^{1/2}  ~.
\end{equation}

What can we say about the probability ${\rm Prob}(H,V)$ that a particular connected path with $(H,V)$ horizontal and vertical links is contained in $E+E_{\rm min}$?
There are altogether $2^{H+V}$ ways to distribute errors (links contained in $E$) at locations on the
specified chain --- each link either has an error or not. And once the error
locations are specified, the probability for errors to occur at those particular
locations is
\begin{eqnarray}
&&p^{H_e}(1-p)^{H-H_e}q^{V_e}(1-q)^{V-V_e}\nonumber\\
&=& (1-p)^H (1-q)^V \left({p\over 1-p}\right)^{H_e}\left({q\over 1-q}\right)^{V_e}~.
\end{eqnarray}
But with those chosen error locations, the cycle can be in $E+E_{\rm min}$ only
if eq.~(\ref{e_ineq}) is satisfied.  Combining these observations, we conclude
that
\begin{equation}
{\rm Prob}(H,V)\le 2^{H+V}\left(\tilde p^{H}\tilde q^{V}\right)^{1/2}~,
\end{equation}
where
\begin{equation}
\tilde p = p(1-p)~,\quad \tilde q=q(1-q)~.
\end{equation}

We can now bound the probability that $E+E_{\rm min}$ contains
any connected path  with $(H,V)$ links (whether an open path or a cycle) by counting such paths. We may think of the path as a walk on the lattice (in the case of a cycle we randomly choose a point on the cycle where the walk begins and ends). Actually, our primary
interest is not in how long the walk is (how many links it contains), but
rather in how far it wanders --- in particular we are interested in whether a closed walk is homologically nontrivial. The walks associated with connected chains of errors visit any given
{\em link} at most once, but it will suffice to restrict the walks further, to
be {\em self-avoiding walks} (SAW's) --- those that visit any given {\em site}
at most once (or in the case of a cycle, revisit only the point where the walk starts and ends).  This restriction proves adequate for our purposes, because given
any open error walk that connects two sites, we can always obtain an SAW by
eliminating some closed loops of links from that walk.  Similarly, given any
homologically nontrivial closed walk, we can obtain a closed SAW (a {\em self-avoiding polygon}, or SAP) by eliminating some links.

If we wish to consider the probability of an error per unit time in the encoded state, we may confine our attention to SAW's that lie between two time slices separated by the finite time $T$. (In fact, we will explain in Sec.~\ref{sec:finite_time} why we can safely assume that $T=O(L)$.)
Such an SAW can begin at any one of $L^2\cdot T$ lattice sites of our
three-dimensional lattice (and in the case of an SAP, we may arbitrarily select one
site that it visits as its ``starting point.'') If $n_{\rm SAP}(H,V)$ denotes
the number of SAP's with $(H,V)$ links and a specified starting site, then the
probability ${\rm Prob}_{\rm SAP}(H,V)$ that $E+E_{\rm min}$ contains any SAP with $(H,V)$ links satisfies
\begin{equation}
\label{saw_prob}
{\rm Prob}_{\rm SAP}(H,V)\le L^2 T \cdot n_{\rm SAP}(H,V)\cdot
2^{H+V}\left(\tilde p^{H}\tilde q^{V}\right)^{1/2}~.
\end{equation}
The upper bound eq.~(\ref{saw_prob}) will be the foundation of the
results that follow.

The encoded quantum information is damaged if $E+E_{\rm min}$ contains homologically nontrivial paths.  At a minimum, the
homologically nontrivial (self-avoiding) path must contain at least $L$
horizontal links.  Hence we can bound the failure probability as
\begin{eqnarray}
\label{saw_L}
& &{\rm Prob}_{\rm fail} \le \sum_{V}\sum_{H\ge L} {\rm Prob}_{\rm
SAP}(H,V)\nonumber\\
& \le & L^2 T \sum_{V}\sum_{H\ge L} n_{\rm SAP}(H,V)
\cdot(4\tilde p)^{H/2} (4\tilde q)^{V/2}~.
\end{eqnarray}

\subsection{Counting anisotropic self-avoiding walks}
\label{subsec:count_walks}

We will obtain  bounds on the accuracy threshold for reliable quantum storage
with toric codes by establishing conditions under which the upper bound
eq.~(\ref{saw_L}) rapidly approaches zero as $L$ gets large. For this analysis, we will need bounds on the number of
self-avoiding polygons with a specified number of horizontal and vertical links. 

One such bound is obtained if we ignore the distinction between horizontal and
vertical links. The first step of an SAP on a simple (hyper)cubic lattice in
$d$ dimensions can be chosen in any of $2d$ directions, and each subsequent
step in at most $2d-1$ directions, so for walks containing a total of $\ell$ links we obtain
\begin{equation}
n_{\rm SAP}^{(d)}(\ell)\le 2d(2d-1)^{\ell-1}~,\quad d~~{\rm dimensions}~.
\end{equation}
Some tighter bounds are known \cite{vanderzande,madras} in the cases $d=2,3$:
\begin{equation}
\label{saw_2}
n_{\rm SAP}^{(2)}(\ell)\le P_2(\ell)(\mu_2)^\ell~,\quad \mu_2\approx 2.638~,
\end{equation}
and
\begin{equation}
\label{saw_3}
n_{\rm SAP}^{(3)}(\ell)\le P_3(\ell)(\mu_3)^\ell~,\quad \mu_3\approx 4.684~,
\end{equation}
where $P_{2,3}(\ell)$ are polynomials.

Since an SAP with $H$ horizontal and $V$ vertical links has $\ell=H+V$ total
links, we may invoke eq.~(\ref{saw_3}) together with eq.~(\ref{saw_L}) to
obtain
\begin{eqnarray}
&&{\rm Prob}_{\rm fail}\nonumber\\
&&\le L^2 T \sum_{V}\sum_{H\ge L} P_3(H+V) \cdot(4\mu_3^2 ~\tilde p)^{H/2}
(4\mu_3^2 ~\tilde q)^{V/2}.
\end{eqnarray}
Provided that
\begin{equation}
\label{threshold_iso}
\tilde p< (4\mu_3^2)^{-1}~,\quad \tilde q < (4\mu_3^2)^{-1}~,
\end{equation}
we have
\begin{equation}
(4\mu_3^2 ~ \tilde p)^{H/2}\cdot (4\mu_3^2 ~ \tilde q)^{V/2} \le (4\mu_3^2 ~\tilde p)^{L/2}~,
\end{equation}
for every term appearing in the sum. Since there are altogether $2L^2 T$
horizontal links and $L^2 T$ vertical links on the lattice, the sum over
${H,V}$ surely can have at most $2L^4T^2$ terms, so that
\begin{equation}
\label{fail_iso}
{\rm Prob}_{\rm fail} < Q_3(L,T)\cdot (4\mu_3^2 ~\tilde p)^{L/2}
\end{equation}
where $Q_3(L,T)$ is a polynomial. 
To ensure that quantum information can be stored with
arbitrarily good reliability, it will suffice that ${\rm Prob}_{\rm fail}$ becomes arbitrarily small as $L$ gets large (with $T$ increasing no faster than a polynomial of $L$).
Thus  eq.~({\ref{threshold_iso}}) is sufficient for reliable quantum storage.
Numerically, the accuracy threshold is surely attained provided that
\begin{equation}
\tilde p, \tilde q < (87.8)^{-1}= .0113~,
\end{equation}
or
\begin{equation}
\label{threshold_iso_num}
p, q <  .0114~.
\end{equation}
Not only does eq.~(\ref{fail_iso}) establish a lower bound on the accuracy threshold; it also shows that, below threshold, the failure probability decreases exponentially with $L$, the square root of the block size of the surface code. 

Eq.~(\ref{threshold_iso_num}) bounds the accuracy threshold in
the case $p=q$, where the sum in eq.~(\ref{saw_L}) is dominated
by isotropic walks with $V\sim H/2$.  But for $q<.0114$, higher values of $p$
can be tolerated, and for $q>.0114$, there is still a threshold, but the
condition on $p$ is more stringent.  To obtain stronger results than
eq.~(\ref{threshold_iso_num}) from eq.~(\ref{saw_L}), we need
better ways to count anisotropic walks, with a specified ratio of $V$ to $H$.

One other easy case is the $q\to 0$ limit (perfect syndrome measurement), where
the only walks that contribute are two-dimensional SAP's confined to a single
time slice. Then we have
\begin{equation}
{\rm Prob}_{\rm fail} < Q_2(L,T)\cdot (4~\mu_2^2 ~\tilde p)^{L/2}
\end{equation}
(where $Q_2(L,T)$ is a polynomial) provided that
\begin{equation}
\tilde p = p(1-p) < (4\mu_2^2)^{-1}\approx (27.8)^{-1}=.0359~,
\end{equation}
or 
\begin{equation}
\label{threshold_iso_2}
p < .0373~;
\end{equation}
the threshold value of $p$ can be relaxed to at least $.0373$ in the case where
syndrome measurements are always accurate. 

This estimate of $p_c$ is considerably smaller than the value $p_c\simeq .1094 \pm .0002$ quoted in Sec.~\ref{subsec:acc_thresh}, obtained from the critical behavior of the random-bond Ising model. That discrepancy is not a surprise, considering the crudeness of our arguments in this section. If one accepts the results of the numerical studies of the random-bond Ising model, and Nishimori's argument that the phase boundary of the model is vertical, then apparently constructing the minimum weight chain is a more effective procedure than our bound indicates.

One possible way to treat the case $q\ne p$ would be to exploit an
observation due to de Gennes \cite{degennes}, which relates the counting of
SAP's to the partition function of a classical $O(N)$ spin model in the limit
$N\to 0$. This spin model is anisotropic, with nearest-neighbor couplings $J_H$
on horizontal links and $J_V$ on vertical links, and its (suitably rescaled) free energy density has the high-temperature expansion
\begin{equation}
\label{partition}
f(J_H,J_V)=\sum_{H,V} n_{\rm SAP}(H,V) \left(J_H\right)^H\left(J_V\right)^V~.
\end{equation}
This expansion converges in the disordered phase of the spin system, but
diverges in the magnetically ordered phase.  Thus, the phase boundary of the
spin system in the $J_H$--$J_V$ plane can be translated into an upper bound on
the storage accuracy threshold in the $p$--$q$ plane, through the relations
\begin{equation}
\tilde p=J^2_H/4~,\quad \tilde q=J^2_V/4~,
\end{equation}
obtained by comparing eq.~(\ref{partition}) and eq.~(\ref{saw_L}).

To bound the failure probability for a planar code rather than the toric code, we should count the ``relative polygons'' that stretch from one edge of the lattice to the opposite edge. This change has no effect on the estimate of the threshold.

\section{Error correction for a finite time interval}
\label{sec:finite_time}

In estimating the threshold for reliable {\em storage} of encoded quantum information, we have found it convenient to imagine that we perform error syndrome measurement forever, without any beginning or end. Thus $S+E$ is a cycle (where $S$ is the syndrome chain and $E$ is the error chain) containing the closed world lines of the defects. Though some of these world lines may be homologically nontrivial, resulting in damage to the encoded qubits, we can recover from the damage successfully if the chain $S+E'$ (where $E'$ is our estimated error chain) is homologically equivalent to $S+E$. The analysis is simplified because we need to consider only the errors that have arisen during preceding rounds of syndrome measurement, and need not consider any pre-existing errors that were present when the round of error correction began. 

However, if we wish to perform a {\em computation} acting on encoded toric
blocks, life will not be so simple.  In our analysis of the storage threshold, we have assumed that the complete syndrome history of an encoded block is known. But when two blocks interact with one another
in the execution of a quantum gate, the defects in each block may propagate to
the other block. Then to assemble a complete history of the defects in any
given block, we would need to take into account the measured syndrome of all
the blocks in the ``causal past'' of the block in question.  In principle this is possible. But in practice, the required classical computation would be far too complex to perform efficiently --- in $T$ parallelized time steps, with two-qubit gates acting in each step, it is
conceivable that defects from as many as $2^T$ different blocks could propagate
to a given block.  Hence, if we wish to compute fault-tolerantly using toric codes, we will need to intervene and perform recovery repeatedly. Since the syndrome measurement is imperfect and the defect positions cannot be precisely determined, errors left over from one round of error correction may cause problems in subsequent rounds.

Intuitively, it should not be necessary to store syndrome information for a very long period to recover successfully, because correlations decay exponentially with time in our statistical-mechanical model. To take advantage of this property, we must modify our recovery procedure.

\subsection{Minimal-weight chains}
Consider performing syndrome measurement $T$ times in succession (starting at time $t=0$), generating syndrome chain $S$ and error chain $E$. Let the error chain $E$ contain any qubit errors that were already present when the syndrome measurements began. Then the chain $S+E$ consisting of all defect world lines contains both closed loops and open paths that end on the final time slice --- we say that $S+E$ is closed relative to the final time slice, or $\partial_{\rm rel}(S+E)=0$. The open connected paths contained in $S+E$ are of two types: pairs of defects created prior to $t=0$ that have persisted until $t=T$ (if the world line contains links on the initial time slice), and pairs of defects created after $t=0$ that have persisted until $t=T$ (if the world line contains no links on the initial slice).

The syndrome $S$ could have been caused by any error chain $E'$ with the same {\em relative} boundary as $E$. To reconstruct the world lines, we should choose an $E'$ that is likely given the observed $S$. A reasonable procedure is to choose the chain $E'$ with $\partial_{\rm rel}E'=\partial_{\rm rel} S$ that minimizes the weight eq.~(\ref{eq:weight}).

The chain $S+E'$ can be projected onto the final time slice --- the projected chain $\Pi(S+E')$ contains those and only those horizonal links that are contained in $S+E'$ on an odd number of time slices. Of course, $E'$ has the same projection as $S+E'$; the syndrome chain $S$ contains only vertical links so that its projection is trivial. The projection $\Pi(E')$ is our hypothesis about which links have errors on the final time slice. After $\Pi(E')$ is constructed, we may perform $X$'s or $Z$'s on these links to compensate for the presumed damage. Note that, to construct $E'$, we do not need to store all of $S$ in our (classical) memory --- only the relative boundary of $S$ is needed.

Actually, any homologically trivial closed loops in $\Pi(E')$ are harmless and can be safely ignored.  Each homologically nontrivial world line modifies the encoded information by the
logical operation $\bar X$ or $\bar Z$.  Thus, after the hypothetical closed
world lines are reconstructed, we may compensate for the homologically
nontrivial closed loops by applying $\bar X$ and/or $\bar Z$ as needed.
Projecting the open world lines in $E'$ onto the final time slice
produces a pairing of the presumed positions of surviving defects on the final
slice.  These defects are removed by performing $Z$'s or $X$'s along a path
connecting the pair that is homologically equivalent to the projected chain
that connects them.  Thus, this recovery step in effect brings the paired
defects together to annihilate harmlessly.

Of course, our hypothesis $E'$ won't necessarily agree exactly with the actual error chain $E$. Thus $E+E'$ contains open chains bounded by the final time slice. Where these open chains meet the final time slice, defects remain that our recovery procedure has failed to remove.

\begin{figure}
\begin{center}
\leavevmode
\epsfxsize=3.5in
\epsfbox{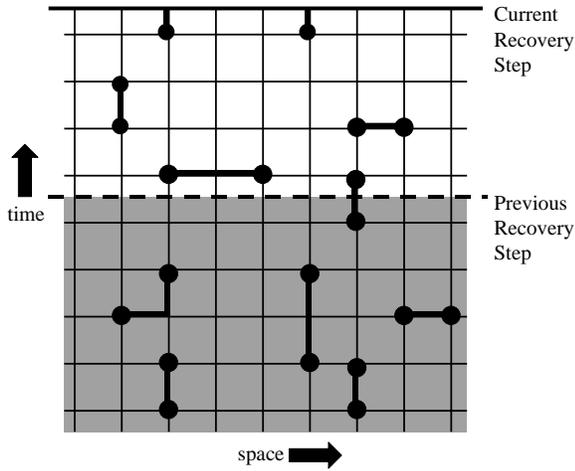}
\end{center}
\caption{The ``overlapping recovery'' method, shown schematically. All monopoles (boundary points of the error syndrome chain) are indicated as filled circles, including both monopoles left over from earlier rounds of error recovery (those in the shaded region below the dotted line) and monopoles generated after the previous round (those in the unshaded region above the dotted line). Also shown is the minimum weight chain $E'$ that connects each monopole to either another monopole or to the current time slice. The chain $E'$ contains $E'_{\rm old}$, whose boundary lies entirely in the shaded region, and the remainder $E'_{\rm keep}$. In the current recovery step, errors are corrected on the horizontal links of $E'_{\rm old}$, and its boundary is then erased from the recorded syndrome history. The boundary of $E'_{\rm keep}$ is retained in the record, to be dealt with in a future recovery step.}
\label{fig:overlap}
\end{figure}

\subsection{Overlapping recovery method}
The procedure of constructing the minimal-weight chain $E'$ with the same {\em relative} boundary as $S$ is not as effective as the procedure in which we continue to measure the syndrome forever. In the latter case, we are in effect blessed with additional information about where monopoles will appear in the future, at times later than $T$, and that additional information allows us to make a more accurate hypothesis about the defect world lines. However, we can do nearly as well if we use a procedure that stores the syndrome history for only a finite time, if we recognize that the older syndrome is more trustworthy than the more recent syndrome. In our statistical physics model, the fluctuating closed loops in $E+E'$ do not grow indefinitely large in either space or in time. Therefore, we can reconstruct an $E'$ that is homologically equivalent to $E$ {\em quasilocally in time} --- to pair up the monopoles in the vicinity of a given time slice, we do not need to know the error syndrome at times that are much earlier or much later.

So, for example, imagine measuring the syndrome $2T$ times in succession (starting at time $t=0$), and then constructing $E'$ with the same relative boundary as $S$. The chain $E'$ can be split into two disjoint subchains, as indicated in Fig.~\ref{fig:overlap}. The first part consists of all connected chains that terminate on two monopoles, where  both monopoles lie in the time interval $0\le t < T$; call this part $E'_{\rm old}$. The rest of $E'$ we call $E'_{\rm keep}$. To recover, we flip the links in the projection $\Pi(E'_{\rm old})$, after which we may erase from memory our record of the monopoles connected by $E'_{\rm old}$; only $E'_{\rm keep}$ (indeed only the relative boundary of $E'_{\rm keep}$) will be needed to perform the next recovery step.

In the next step we measure the syndrome another $T$ times in succession, from $t=2T$ to $t=3T-1$. Then we choose our new $E'$ to be the minimal-weight chain whose boundary relative to the new final time slice is the union of the relative boundary of $S$ in the interval $2T \le t < 3T$ and the relative boundary of $E'_{\rm keep}$ left over from previous rounds of error correction. We call this procedure the ``overlapping recovery method'' because the minimal-weight chains that are constructed in successive steps occupy overlapping regions of spacetime.

If we choose $T$ to be large compared to the characteristic correlation time of our statistical physics model, then only rarely will a monopole survive for more than one round, and the amount of syndrome information we need to store will surely be bounded. Furthermore, for such $T$, this overlapping recovery method will perform very nearly as well as if an indefinite amount of information were stored.

The time $T$ should be chosen large enough so that connected chains in $E+E'$ are not likely to extend more than a distance $T$ in the time direction. Arguing as in Sec.~\ref{subsec:count_walks} (and recalling that the number $n_{\rm SAW}(\ell)$ of self-avoiding walks of length $\ell$ differs from the number $n_{\rm SAP}(\ell)$ of self-avoiding polygons of length $\ell$ by a factor polynomial in $\ell$), we see that a connected chain containing $H$ horizontal links and $V$ vertical links occurs with a probability
\begin{equation}
{\rm Prob}(H,V) \le Q_3'(H,V) (4\mu_3^2\tilde p)^{H/2}(4\mu_3^2\tilde q)^{V/2}~,
\end{equation}
where $Q_3'(H,V)$ is a polynomial.
Furthermore, a connected chain with temporal extent $T$ must have at least $V=2T$ vertical links if both ends of the chain lie on the final time slice. Therefore the probability ${\rm Prob}(H,V)$ is small compared to the failure probability eq.~(\ref{fail_iso}), so that our procedure with finite memory differs in efficacy from the optimal procedure with infinite memory by a negligible amount, provided that
\begin{equation}
T \gg {L\over 2} \cdot {\log(4\mu_3^2\tilde p)^{-1}\over \log(4\mu_3^2\tilde q)^{-1}}~.
\end{equation}
In particular, if the measurement error and qubit error probabilities are comparable ($q\simeq p$), it suffices to choose $T\gg L$, where $L$ is the linear size of the lattice.

Thus we see that the syndrome history need not be stored indefinitely for our recovery procedure to be robust. The key to fault tolerance is that we should not overreact to syndrome information that is potentially faulty. In particular, if we reconstruct the world lines of the defects and find open world lines that do not extend very far into the past, it might be dangerous to accept the accuracy of these world lines and respond by bringing the defects together to annihilate. But world lines that persist for a time comparable to $L$ are likely to be trustworthy. In our overlapping recovery scheme, we take action to remove only these long-lived defects, leaving those of more recent vintage to be dealt with in the next recovery step. 

\subsection{Computation threshold}
\label{subsec:comp_thresh}
Our three-dimensional model describes the history of a single code block; hence its phase transition identifies a threshold for reliable {\rm storage} of quantum information. Analyzing the threshold for reliable quantum {\em computation} is more complex, because we need to consider interactions between code blocks.

When two encoded blocks interact through the execution of a gate, errors can propagate from one block to another, or potentially from one qubit in a block to another qubit in the same block. It is important to keep this error propagation under control. We will discuss in Sec.~\ref{sec:ftgates} how a universal set of fault-tolerant quantum gates can be executed on encoded states. For now let us consider the problem of performing a circuit consisting of CNOT gates acting on pairs of encoded qubits. The encoded CNOT gate with block 1 as its control and block 2 as its target can be implemented {\em transversally} --- that is, by performing CNOT gates in parallel, each acting on a qubit in block 1 and the corresponding qubit in block 2. A CNOT gate propagates bit-flip errors from control to target and phase errors from target to control. Let us first consider the case in which storage errors occur at a constant rate, but errors in the gates themselves can be neglected. 

Suppose that a transversal CNOT gate is executed at time $t=0$, propagating bit-flip errors from block 1 to block 2, and imagine that we wish to correct the bit-flip errors in block 2. We suppose that many rounds of syndrome measurement are performed in both blocks  before and after $t=0$. Denote by $S_1$ and $S_2$ the syndrome chains in the two blocks, and by $E_1$ and $E_2$ the error chains. Due to the error propagation, the chain $S_2+E_2$ in block 2 has a nontrivial boundary at the $t=0$ time slice. Therefore, to diagnose the errors in block 2 we need to modify our procedure.  

We may divide each syndrome chain and error chain into two parts, a portion lying in the past of the $t=0$ time slice, and a portion lying in its future. Then the chain 
\begin{eqnarray}
&&S_{1,\rm before} + S_{2,\rm before} +S_{2,\rm after} \nonumber\\
&&+E_{1,\rm before} + E_{2,\rm before} +E_{2,\rm after}~
\end{eqnarray}
has a trivial boundary. Therefore, we can estimate $E_{1,\rm before} + E_{2,\rm before} +E_{2,\rm after}$ by constructing the minimal chain with the same boundary as $S_{1,\rm before} + S_{2,\rm before} +S_{2,\rm after}$. Furthermore, because of the error propagation, it is $E_{1,\rm before} + E_{2,\rm before} +E_{2,\rm after}$ whose horizontal projection identifies the damaged links in block 2 after $t=0$. 

If in each block the probability of error per qubit and per time step is $p$, while the probability of a syndrome measurement error is $q$, then the error chain $E_{1,\rm before} + E_{2,\rm before} +E_{2,\rm after}$ has in effect been selected from a distribution in which the error probabilities are $(2p(1-p),2q(1-q))$ before the gate, and $(p,q)$ after the gate. Obviously, these errors are no more damaging than if the error probabilities had been $(2p(1-p),2q(1-q))$ at all times, both before and after $t=0$. Therefore, if $(p,q)$ lies below the accuracy threshold for accurate storage, then error rates $(2p(1-p),2q(1-q))$ will be below the accuracy threshold for a circuit of CNOT gates.

Of course, the transversal CNOT might itself be prone to error, damaging each qubit with probability $p_{\rm CNOT}$, so that the probability of error is larger on the $t=0$ slice than on earlier or later slices. However, increasing the error probability from $p$ to $p+p_{\rm CNOT}$ on a single slice is surely no worse than increasing the probability of error to $p+p_{\rm CNOT}$ on all slices. For a given $q$, there is a threshold value $p_c(q)$, such that for $p<p_c(q)$ a circuit of CNOT's is robust if the gates are flawless; then the circuit with imperfect gates is robust provided that $p+p_{\rm CNOT} < p_c(q)$.

By such reasoning, we can infer that the accuracy threshold for quantum computation is comparable to the threshold for reliable storage, differing by factors of order one. Furthermore, below threshold, the probability of error in an encoded gate decreases exponentially with $L$, the linear size of the lattice. Therefore, to execute a quantum circuit that contains $T$ gates with reasonable fidelity, we should choose $L=O(\log T)$, so that the block size $2L^2$ of the code is $O(\log^2 T)$.

\section{Quantum circuits for syndrome measurement}
\label{sec:storage_gates}

In our model with uncorrelated errors, in which qubit errors occur with probability $p$ per time step and measurement errors occur with probability $q$, we have seen in Sec.~\ref{sec:statphys} that it is possible to identify a sharp phase boundary between values of the parameters such that error correction is sure to succeed in the limit of a large code block, and values for which error correction need not succeed. How can we translate this accuracy threshold, expressed as a phase boundary in the $p$--$q$ plane, into a statement about how well the hardware in our quantum memory must perform in order to protect quantum states effectively? The answer really depends on many details about the kinds of hardware that are potentially at our disposal. For purposes of illustration, we will relate $p$ and $q$ to the error probabilities for the fundamental gates in a particular computational model. 

\subsection{Syndrome measurement}
Whenever a check operator $X_s$ or $Z_P$ is measured, a quantum circuit is executed in which each of the qubits occuring in the check operator interacts with an ancilla, and then the ancilla is measured to determine the result. Our task is to study this quantum circuit to determine how the faults in the circuit contribute to $p$ and to $q$. To start we must decide what circuit to study. 

For many quantum codes, the design of the syndrome measurement circuit involves subtleties. If the circuit is badly designed, a single error in the ancilla can propagate to many qubits in the code block, compromising the effectiveness of the error correction procedure. To evade this problem, Shor \cite{shor_ft} and Steane \cite{steane_anc} proposed two different methods for limiting the propagation of error from ancilla to data in the measurement of the check operators of a stabilizer code. In Shor's method, to extract each bit of the error syndrome, an ancilla ``cat state'' is prepared that contains as many qubits as the weight of the check operator. The ancilla interacts with the data code block, and then each qubit of the ancilla is measured; the value of the check operator is the parity of the measurement outcomes. In Steane's method, the ancilla is prepared as an encoded block (containing as many qubits as the length of the code). The ancilla interacts with the data, each qubit in the ancilla is measured, and a classical parity check matrix is applied to the measurement outcomes to extract the syndrome. In either scheme, each ancilla qubit interacts with only a single qubit in the data, so that errors in the ancilla cannot seriously damage the data. The price we pay is the overhead involved in preparing the ancilla states and verifying that the preparation is correct.

We could use the Shor method or the Steane method to measure the stabilizer of a surface code, but it is best not to. We can protect against errors more effectively by using just a single ancilla qubit for the measurement of each check operator, avoiding all the trouble of preparing and verifying ancilla states. The price we pay is modest --- a single error in the ancilla might propagate to become two errors in the data, but we'll see that these correlated errors in the data are not so damaging. 

So we imagine placing a sheet of ancilla qubits above the qubits of a planar code block. Directly above the site $s$ is the ancilla qubit that will be used to measure $X_s$, and directly above the center of the plaquette $P$ is the ancilla qubit that will be used to measure $Z_P$. We suppose that CNOT gates can be executed acting on a data qubit and its neighboring ancilla qubits. The circuits for measuring the plaquette operator $Z^{\otimes 4}$ and the site operator
$X^{\otimes 4}$ are shown in Fig.~\ref{fig:nft}:

\begin{figure}[h]
\centering
\begin{picture}(230,110)

\put(0,40){\makebox(0,0){data}}
\put(0,60){\makebox(0,0){data}}
\put(0,80){\makebox(0,0){data}}
\put(0,100){\makebox(0,0){data}}

\put(10,10){\line(1,0){65}}
\put(10,40){\line(1,0){65}}
\put(10,60){\line(1,0){65}}
\put(10,80){\line(1,0){65}}
\put(10,100){\line(1,0){65}}

\put(20,100){\circle*{4}}
\put(20,100){\line(0,-1){94}}
\put(20,10){\circle{8}}

\put(35,80){\circle*{4}}
\put(35,80){\line(0,-1){74}}
\put(35,10){\circle{8}}

\put(50,60){\circle*{4}}
\put(50,60){\line(0,-1){54}}
\put(50,10){\circle{8}}

\put(65,40){\circle*{4}}
\put(65,40){\line(0,-1){34}}
\put(65,10){\circle{8}}

\put(85,4){\makebox(12,12){meas.}}

\put(110,40){\makebox(0,0){data}}
\put(110,60){\makebox(0,0){data}}
\put(110,80){\makebox(0,0){data}}
\put(110,100){\makebox(0,0){data}}

\put(110,10){\line(1,0){5}}
\put(120,40){\line(1,0){82}}
\put(120,60){\line(1,0){82}}
\put(120,80){\line(1,0){82}}
\put(120,100){\line(1,0){82}}

\put(115,4){\framebox(12,12){$H$}}
\put(127,10){\line(1,0){65}}

\put(137,10){\circle*{4}}
\put(137,10){\line(0,1){94}}
\put(137,100){\circle{8}}

\put(152,10){\circle*{4}}
\put(152,10){\line(0,1){74}}
\put(152,80){\circle{8}}

\put(167,10){\circle*{4}}
\put(167,10){\line(0,1){54}}
\put(167,60){\circle{8}}

\put(182,10){\circle*{4}}
\put(182,10){\line(0,1){34}}
\put(182,40){\circle{8}}

\put(192,4){\framebox(12,12){$H$}}
\put(204,10){\line(1,0){5}}

\put(219,4){\makebox(12,12){meas.}}

\end{picture}
\caption{Circuits for measurement of the plaquette ($Z^{\otimes 4}$)
and site ($X^{\otimes 4}$) stabilizer operators.}
\label{fig:nft}
\end{figure}
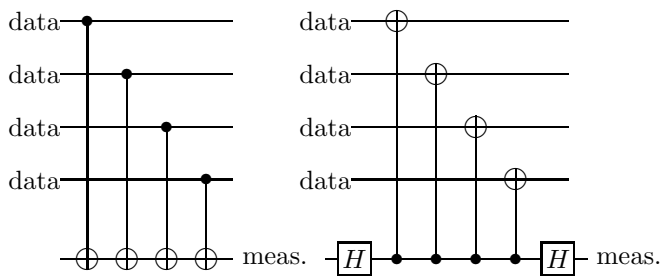

\noindent We have included the Hadamard gates in the circuit for measuring the site operator to signify that the ancilla qubit is initially prepared in the $X=1$ state, and the final measurement is a measurement of $X$, while in the case of the plaquette operator measurement the ancilla is prepared in the $Z=1$ state and $Z$ is measured at the end. But we will suppose that our computer can measure $X$ as easily as it can measure $Z$; hence in both cases the circuit is executed in six time steps (including preparation and measurement), and there is really no Hadamard gate.

\subsection{Syndrome errors and data errors}
We will assume that all errors in the circuit are stochastic (for example, they could be errors caused by decoherence). We will consider both ``storage errors'' and ``gate errors.'' In each time step, the probability that a ``resting'' qubit is damaged will be denoted $p_s$. For simplicity, we will assume that an error, when it occurs, is one of the Pauli operators $X$, $Y$, or $Z$. (The analysis of the circuit is easily generalized to more general models of stochastic errors.) In our analysis, we will always make a maximally pessimistic assumption about which error occured at a particular position in the circuit. If a gate acts on a qubit in a particular time step, we will assume that there is still a probability $p_s$ of a storage error in that step, plus an additional probability of error due to the execution of the gate. We denote the probability of an error in the two-qubit CNOT
gate by $p_{\rm CNOT}$; the error is a tensor product of Pauli operators, and again we will always make maximally pessimistic assumptions about which error occurs at a particular position in the circuit.  If a storage error and gate error occur in the same time step, we assume that the gate error acts first, followed by the storage error. When a single qubit is
measured in the $\{\ket{0},\ket{1}\}$ basis, $p_m$ is the probability of
obtaining the incorrect outcome.  (If a storage error occurs during a measurement step, we assume that the error precedes the measurement.) And when a fresh qubit is acquired in the
state $\ket{0}$, $p_p$ denotes the probability that its preparation is faulty
(it is $\ket{1}$ instead). 

\begin{figure}[h]
\centering
\begin{picture}(230,110)

\put(0,10){\makebox(0,0){$P_{\rm west}$}}
\put(0,30){\makebox(0,0){$s_{\rm north}$}}
\put(0,50){\makebox(0,0){$P_{\rm east}$}}
\put(0,70){\makebox(0,0){$s_{\rm south}$}}
\put(0,100){\makebox(0,0){data}}

\put(15,10){\line(1,0){85}}
\put(15,30){\line(1,0){85}}
\put(15,50){\line(1,0){85}}
\put(15,70){\line(1,0){85}}
\put(15,100){\line(1,0){85}}

\put(35,70){\circle*{4}}
\put(35,70){\line(0,1){34}}
\put(35,100){\circle{8}}

\put(50,100){\circle*{4}}
\put(50,100){\line(0,-1){54}}
\put(50,50){\circle{8}}

\put(65,30){\circle*{4}}
\put(65,30){\line(0,1){74}}
\put(65,100){\circle{8}}

\put(80,100){\circle*{4}}
\put(80,100){\line(0,-1){94}}
\put(80,10){\circle{8}}

\put(130,10){\makebox(0,0){$s_{\rm west}$}}
\put(130,30){\makebox(0,0){$P_{\rm north}$}}
\put(130,50){\makebox(0,0){$s_{\rm east}$}}
\put(130,70){\makebox(0,0){$P_{\rm south}$}}
\put(130,100){\makebox(0,0){data}}

\put(145,10){\line(1,0){85}}
\put(145,30){\line(1,0){85}}
\put(145,50){\line(1,0){85}}
\put(145,70){\line(1,0){85}}
\put(145,100){\line(1,0){85}}

\put(165,100){\circle*{4}}
\put(165,100){\line(0,-1){34}}
\put(165,70){\circle{8}}

\put(180,50){\circle*{4}}
\put(180,50){\line(0,1){54}}
\put(180,100){\circle{8}}

\put(195,100){\circle*{4}}
\put(195,100){\line(0,-1){74}}
\put(195,30){\circle{8}}

\put(210,10){\circle*{4}}
\put(210,10){\line(0,1){94}}
\put(210,100){\circle{8}}

\end{picture}
\caption{Gates acting on a given qubit in a complete round of syndrome measurement. Data qubits on links with a north-south orientation  participate successively in measurements of check operators at the site to the south, the plaquette to the east, the site to the north, and the plaquette to the west. Qubits on links with an east-west orientation  participate successively in measurements of check operators at the plaquette to the south, the site to the east, the plaquette to the north, and the site to the west.}
\label{fig:nft_data}
\end{figure}
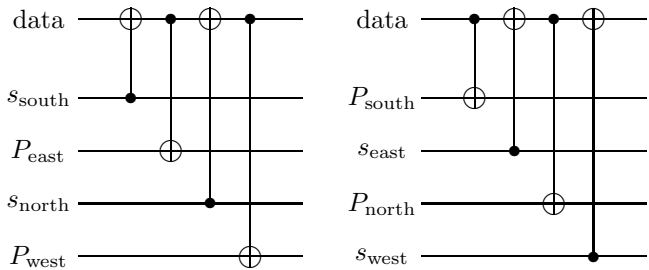

In a single cycle of syndrome measurement, each data qubit participates in the
measurement of four stabilizer operators: two site operators and two plaquette
operators.  Each of these measurements requires four time steps (excluding the preparation and measurement steps), as a single
ancilla qubit is acted upon by four sequential CNOT's.  But to cut down the
likelihood of storage errors, we can execute the four measurement circuits
in parallel, so that every data qubit participates in a CNOT gate in every
step. For example, for each plaquette and each site, we may execute CNOT gates that act on the four edges of the plaquette or the four links meeting at the site in the counterclockwise order north-west-south-east. The CNOT gates that act on a given data qubit, then, alternate between CNOT's with the data qubit as control and CNOT's with the data qubit as target, as indicated in Fig.~\ref{fig:nft_data}.

For either a site check operator or a plaquette check operator, the probability that the measurement is faulty is
\begin{equation}
q_{\rm single} = p_p + 4 p_{\rm CNOT} + 6 p_s + p_m + {\rm h.~o.}~,
\label{q_est}
\end{equation}
where ``$+~{\rm h.~o.}$'' denotes terms of higher than linear order in the fundamental error probabilities. The measurement can fail if any one of the CNOT gates has an error, if a storage error occurs during any of the six time steps needed to execute the circuit (including the preparation and measurement step), or because of a fault in the initial preparation or final measurement of the ancilla qubit. By omitting the higher order terms we are actually {\em overestimating} $q$. For example, $p_s$ is the probability that a storage error occurs in the first time step, disregarding whether or not additional errors occur in the circuit.

We have used the notation $q_{\rm single}$ in eq.~(\ref{q_est}) to emphasize that this is an estimate of the probability of an isolated error on a vertical (timelike) link. 
More troublesome are syndrome measurement errors that are correlated with qubit errors. These arise if, say, a qubit suffers a $Z$ error that is duly recorded in the syndrome measurement of one of the two adjoining sites but not the other. In our spacetime picture, then, there is a timelike plaquette with an error on one of its horizontal links and one of its vertical links. We will refer to this type of correlated error as a ``vertical hook'' --- hook because the two links with errors meet at a $90^\circ$ angle, and vertical because one of the links is vertical (and to contrast with the case of a horizontal hook which we will discuss later). 

We can estimate the probability of a vertical hook on a specified timelike plaquette by considering the circuits in Fig.~\ref{fig:nft_data}. The qubit in question participates in the measurement of two site check operators, through the two CNOT gates in the circuit in which the data qubit is the target of the CNOT. A vertical hook can arise due to a fault that occurs in either of these CNOT gates or at a time in between the execution of these gates. Hence the probability of a vertical hook is
\begin{equation}
q_{\rm hook}=3p_{\rm CNOT} + 2 p_s +{\rm h.~o.}~;
\end{equation}
faults in any of three different CNOT gates, or storage errors in either of two time steps, can generate the hook. Note that the hook on the specified plaquette has a unique orientation; the first of the two site operator measurements that the data qubit participated in is the one that fails to detect the error. Of course, the same formula for $q_{\rm hook}$ applies if we are considering the measurement of the plaquette operators rather than the site operators.

A CNOT gate propagates $X$ errors from control qubit to target qubit, and $Z$ errors from target to control. Thus we don't have to worry about a vertical hook that arises from an error in an ancilla bit that propagates to the data. For example, if we are measuring a plaquette operator, then $X$ errors in the ancilla damage the syndrome bit while $Z$ errors in the ancilla propagate to the data; the result is a vertical error in the $X$-error syndrome that is correlated with a horizontal $Z$-error in the data. This correlation is not problematic because we deal with $X$ errors and $Z$ errors separately. However, propagation of error from ancilla to data also generates correlated horizontal errors that we need to worry about. In the measurement of, say, the plaquette operator $Z_P=Z^{\otimes 4}$, $Z$ errors (but not $X$ errors) can feed back from the ancilla to the data.  Feeding back four $Z$'s
means no error at all, because $Z^{\otimes 4}$ is in the code stabilizer, and feeding back three $Z$'s generates the error $IZZZ$, which is equivalent to the single $Z$ error $ZIII$. Therefore, the only way to get a double qubit error from a single fault in the circuit is through an error in the second or third CNOT, or through an ancilla storage error in between the second and third CNOT. (The second CNOT might apply $Z$ to the ancilla but not to the data, and that $Z$ error in the ancilla can then feed back to two data qubits, or the third CNOT could apply $Z$ to both ancilla and data, and the $Z$ error in the ancilla can then feed back to one other data qubit.) Because of the order we have chosen for the execution of the CNOT's, this double error, when it occurs, afflicts the southeast corner of the plaquette (or equivalently the northwest corner, which has the same boundary). We will refer to this two-qubit error as a ``horizontal hook,'' because the two horizontal errors meet at a $90^\circ$ angle. Similarly, error propagation during the measurement of the site operator $X_s$ can produce $X$ errors on the north and west links meeting at that site. One should emphasize that the only correlated $XX$ or $ZZ$ errors that occur with a probability linear in the fundamental error probabilities are these hooks. This is a blessing --- correlated errors affecting two collinear links would be more damaging.

Feedback from the measurement of a plaquette operator can produce $ZZ$ hooks but not $XX$ hooks, and feedback from the measurement of a site operator can produce $XX$ hooks but not $ZZ$ hooks. Thus, in each round of syndrome measurement, the probability of a $ZZ$ hook at a plaquette or an $XX$ hook at a site is
\begin{equation}
p_{\rm hook} = 2 p_{\rm CNOT} + p_s +{\rm h.~o.}~.
\end{equation}
(Remember that a ``hook'' means two $Z$'s or two $X$'s; in addition, an error in a single CNOT gate could induce, say, an $X$ error in the data and a $Z$
error in the ancilla that subsequently feeds back, but correlated $X$ and $Z$
errors will not cause us any trouble.) 

Now we need to count the ways in which a single error can occur in the data during a round of syndrome measurement. First suppose that we measure a single pla\-quette operator $Z_P$, and consider the scenarios that lead to a single $Z$ error in the data. The $Z$ error can arise either because a gate or storage error damages the data qubit directly, or because an error in the ancilla feeds back to the data. Actually, single errors occur with slightly different probabilities for different data qubits acted on by the circuit.  The worst case occurs for the first and last qubit acted on by the circuit; the probability that the circuit produces a single error that acts on the first (or last) qubit is 
\begin{eqnarray}
p_{\rm single,Z}^{Z_P,1}&=&p_{\rm single,Z}^{Z_P,4}\nonumber\\
&=&p_{\rm CNOT} + 6p_s~  + p_{\rm
CNOT} + p_s + {\rm h.~o.}~.
\end{eqnarray}
The first two terms arise from gate errors and storage errors that damage the data qubit directly. For the first qubit, the last two terms arise from the case in which a $Z$ error in the ancilla is fed back to the data by each of the last three CNOT's --- the resulting $IZZZ$ error is equivalent to a $ZIII$ error because $ZZZZ$ is in the code stabilizer. For the fourth qubit, the last two terms arise from an error fed back by the last CNOT gate in the circuit. On the other hand, for the second and third qubit acted on by the circuit, it isn't possible for just a single error to feed back; {\it e.g.}, if the error feeds back to the third qubit, it will feed back to the fourth as well, and the result will be a hook instead of a single error. Hence, the probability of a single error acting on the second or third qubit is\begin{equation}
p_{\rm single,Z}^{Z_P,2}=p_{\rm single,Z}^{Z_P,3}=p_{\rm CNOT} + 6p_s+{\rm h.~o.}
~;
\end{equation}
there is no feedback term.
If we are measuring a site operator $X_s$, then $X$ errors might feed back from the ancilla to the data, but $Z$ errors will not. Therefore, for each of the four qubits acted on by the circuit, the probability that a single $Z$ error results from the execution of the circuit, acting on that particular qubit, is
\begin{equation}
p_{\rm single,Z}^{X_s}=p_{\rm CNOT} + 6p_s+{\rm h.~o.}
~;
\end{equation}
again there is no feedback term.

In a single round of syndrome measurement, each qubit participates in the measurement of four check operators, two site operators and two plaquette operators. For the plaquette operator measurements, depending on the orientation of the link where the qubit resides, the qubit will be either the first qubit in one measurement and the third in the other, or the second in one and the fourth in the other. Either way, the total probability of a single $Z$ error arising that afflicts that qubit is
\begin{eqnarray}
p_{\rm single}&= &4 p_{\rm CNOT} + 6 p_s + p_{\rm CNOT} + p_s+{\rm
h.~o.}\nonumber\\
&=& 5p_{\rm CNOT} + 7 p_s +{\rm h.~o.}~,
\end{eqnarray}
with the $4 p_{\rm CNOT} + 6 p_s$ arising from direct damage to the qubit and the $p_{\rm CNOT} + p_s$ from feedback due to one of the four check operator measurements. The same equation applies to the probability of a single $X$ error arising at a given qubit in a single round of syndrome measurement.

\subsection{Error-chain combinatorics}

With both single errors and hooks to contend with, it is more
complicated to estimate the failure probability, but we can still obtain useful upper
bounds.  In fact, the hooks don't modify the estimate of the accuracy threshold as much as might have been naively expected. Encoded information is damaged if $E+E_{\rm min}$ contains a homologically nontrivial (relative) cycle, which can wrap around the code block with either a north-south or east-west orientation. Either way, the cycle contains at least $L$ links all with the {\em same} orientation, where $L$ is the linear size of the lattice. A horizontal hook introduces two errors with {\em different} orientations, which is not as bad as two errors with the same orientation. Similarly, a vertical hook contains only one horizontal error.

There are two other reasons why the hooks do not badly compromise the effectiveness of error correction. While single errors can occur with any orientation, horizontal hooks can appear only on the northwest corner of a plaquette (hooks on southeast corners are equivalent to hooks on northwest corners and should not be counted separately), and vertical hooks on timelike plaquettes have a unique orientation, too. Therefore, hooks have lower ``orientational entropy'' than the single errors, which means that placing hooks on self-avoiding walks reduces the number of walks of a specified length. And finally, $p_{\rm hook}$ is smaller than $p_{\rm single}$, and $q_{\rm hook}$ is smaller than $q_{\rm single}$, which further reduces the incentive to include hooks in $E+E_{\rm min}$.

We will suppose that $E_{\rm min}$ is constructed by the same procedure as before, by minimizing the weight
\begin{equation}
H\log p_{\rm single}^{-1} + V\log q_{\rm single}^{-1}~.
\end{equation}
To simplify later expressions, we have replaced $p/(1-p)$ by $p$ here, which will weaken our upper bound on the failure probability by an insignificant amount. Note that our procedure finds the most probable chain under the assumption that only single errors occur (no hooks). If $p_{\rm hook}$ and $q_{\rm hook}$ are assumed to be known, then in principle we could retool our
recovery procedure by taking these correlated errors into account in the
construction of $E_{\rm min}$.  To keep things simple we won't attempt to do that.
Then, as before, for
any connected subchain of $E+E_{\rm min}$ with $H$ horizontal links and $V$
vertical links, the numbers $H_e$ and $V_e$ of horizontal and vertical links of
the subchain that are contained in $E$ must satisfy
\begin{equation}
p_{\rm single}^{H_e}q_{\rm single}^{V_e}\le p_{\rm single}^{H/2} q_{\rm single}^{V/2}~.
\end{equation}

To bound the failure probability, we wish to count the number of ways in which a connected chain with a specified number of horizontal links can occur in $E+E_{\rm min}$, keeping in mind that  the error chain $E$ could contain hooks as well as single errors. Notice that a hook might contribute only a single link to $E+E_{\rm min}$, if one of the links contained in the hook is also in $E_{\rm min}$. But since $p_{\rm hook}<p_{\rm single}$ and $q_{\rm hook}<q_{\rm single}$, we will obtain an upper bound on the failure probability if we pessimistically assume  that all of the errors in $E+E_{\rm min}$ are either two-link hooks occuring with probabilities $p_{\rm hook},q_{\rm hook}$ or single errors occuring with probabilities $p_{\rm single},q_{\rm single}$. If the $H_e$ horizontal errors on a connected chain include $H_{\rm hook}$ horizontal hooks and $V_{\rm hook}$ vertical hooks, then there are $H_e-2H_{\rm hook}- V_{\rm hook}$ single horizontal errors and $V_e-V_{\rm hook}$ single vertical errors; once the locations of the hooks and the single errors are specified, the probability that errors occur at those locations is no larger than
\begin{eqnarray}
&&(p_{\rm single})^{H_e-2H_{\rm hook}-V_{\rm hook}}(p_{\rm hook})^{H_{\rm
hook}}\nonumber\\
&&\quad\cdot ~(q_{\rm single})^{V_e-V_{\rm hook}}(q_{\rm hook})^{V_{\rm hook}}\nonumber\\
&&<p_{\rm single}^{H/2}\left({p_{\rm hook}\over p_{\rm single}^2}\right)^{H_{\rm hook}}q_{\rm single}^{V/2}\left({q_{\rm hook}\over p_{\rm single} q_{\rm single}}\right)^{V_{\rm hook}}~.
\label{location_bound}
\end{eqnarray}

Because a horizontal hook contains two errors with different orientations, it will be convenient to distinguish between links oriented east-west and links oriented north-south. We denote by $H_1$ the number of horizontal links in the connected chain with east-west orientation and by $H_2$ the number of horizontal links with north-south orientation; then clearly 
\begin{equation}
H_{\rm hook}\le H_1~,\quad H_{\rm hook} \le H_2~.
\end{equation}
To estimate the threshold, we will bound the probability that our connected chain has $H_1\ge L$; of course, the same expression bounds the probability that $H_2\ge L$.

For a specified connected chain, suppose that altogether $H_e$ of the horizontal links and $V_e$ of the vertical links have errors, and that there are $H_{\rm hook}$ horizontal hooks and $V_{\rm hook}$ vertical hooks, so that there are $H_e-2H_{\rm hook}-V_{\rm hook}$ single horizontal errors and $V_e-V_{\rm hook}$ single vertical errors. In how many ways can we distribute the hooks and single errors along the path? Since each horizontal hook contains a link with north-south orientation, there are no more than ${H_2}\choose{H_{\rm hook}}$ ways to choose the locations of the horizontal hooks; similarly there are no more than ${V\choose V_{\rm hook}}$ ways to choose the locations of the vertical hooks.\footnote{Actually, we have given short shrift here to a slight subtlety. Once we have decided that a vertical hook will cover a particular vertical link, there may be two ways to place the hook --- it might cover either one of two adjacent horizontal links. However, for the hook to be free to occupy either position, the orientation of the second horizontal link must be chosen in one of only two possible ways.  Thus the freedom to place the hook in two ways is more than compensated by the reduction in the orientational freedom of the other horizontal link by a factor of 2/5, and can be ignored. A similar remark applies to horizontal hooks.} Then there are no more than $2^{H_1+H_2-2H_{\rm hook}-V_{\rm hook}}$ ways to place the single horizontal errors among the remaining horizontal links, and no more than $2^{V-V_{\rm hook}}$ ways to place the single vertical errors among remaining $V-V_{\rm hook}$ vertical links on the chain. 
Now consider counting the self-avoiding paths starting at a specified site, where the path is constructed from hooks, single errors, and the links of $E_{\rm min}$. Whenever we add a horizontal hook to the path there are at most two choices for the orientation of the hook, and whenever we add a vertical hook there are at most four choices; hence there are no more than $2^{H_{\rm hook}}4^{V_{\rm hook}}$ ways to choose the orientations of the hooks. For the remaining $H_1 +H_2 - 2H_{\rm hook} +V- 2V_{\rm hook}$ links of the path, the orientation can be chosen in no more than 5 ways. Hence, the total number of paths with a specified number of horizontal links, horizontal hooks, vertical links, and vertical hooks is no more than
\begin{eqnarray}
{{H_2}\choose{H_{\rm hook}} }&&{{V}\choose{V_{\rm hook}} }\cdot 2^{H_1+H_2-2H_{\rm hook}-V_{\rm hook}}2^{V-V_{\rm hook}}\nonumber\\
&& \cdot2^{H_{\rm hook}}4^{V_{\rm hook}}\cdot 5^{H_1 +H_2 - 2H_{\rm hook} +V-2V_{\rm hook}}~.
\end{eqnarray}
Combining this counting of paths with the bound eq.~(\ref{location_bound}) on the probability of each path, we conclude that the  probability that $E+E_{\rm min}$ contains a connected path with specified starting site, containing $H_1$ links with east-west orientation, $H_2$ links with north-south orientation, $V$ vertical links, $H_{\rm hook}$ horizontal hooks, and $V_{\rm hook}$ vertical hooks is bounded above by
\begin{eqnarray}
&&{H_2\choose H_{\rm hook}} \left(p_{\rm hook}\over 50 p_{\rm single}^2\right)^{H_{\rm hook}}\left(100 p_{\rm single}\right)^{(H_1+H_2)/2}\nonumber\\
&&\cdot {{V}\choose{V_{\rm hook}} }\left(q_{\rm hook}\over 25 p_{\rm single}q_{\rm single}\right)^{V_{\rm hook}}
\cdot(100 q_{\rm single})^{V/2}~.
\end{eqnarray}
Here $H_{\rm hook}$ can take any value from zero to $H_2$, and $V_{\rm hook}$ can take any value from zero to $V$. We can sum over $H_{\rm hook}$ and $V_{\rm hook}$, to obtain an upper bound on the probability of a chain with an unspecified number of hooks:
\begin{eqnarray}
&&\left(100 p_{\rm single}\right)^{(H_1+H_2)/2}\left(1 +{p_{\rm hook}\over 50 p_{\rm single}^2}\right)^{H_2}\nonumber\\
&&\cdot (100 q_{\rm single})^{V/2}\left(1 +{q_{\rm hook}\over 25 p_{\rm single}q_{\rm single}}\right)^{V}~.
\end{eqnarray}
Finally, since a path can begin at any of $L^2T$ sites, and since there are two types of homologically nontrivial cycles, the probability of failure ${\rm Prob}_{\rm fail}$ satisfies the bound
\begin{eqnarray}
&&{\rm Prob}_{\rm fail} < 2L^2T \sum_{H_1\ge L} \left(100 p_{\rm single}\right)^{H_1/2}
\nonumber\\
&&\cdot  \sum_{H_2\ge 0}\left[100 p_{\rm single}\left(1 +{p_{\rm hook}\over 50 p_{\rm single}^2}\right)^2\right]^{H_2/2}\nonumber\\
&&\cdot\sum_{V\ge 0} \left[100 q_{\rm single}\left(1 +{q_{\rm hook}\over 25 p_{\rm single}q_{\rm single}}\right)^2\right]^{V/2}
~.
\end{eqnarray}
This sum will be exponentially small for large $L$ provided that
\begin{eqnarray}
&&p_{\rm single}< {1\over 100}~,\quad  q< {1\over 100}~,\nonumber\\
&&p_{\rm hook} < 5~ p_{\rm single}^2\left({1\over \sqrt{p_{\rm single}}}-10\right)~,\nonumber\\
&&q_{\rm hook} < {5\over 2}~ p_{\rm single}q_{\rm single}\left({1\over \sqrt{q_{\rm single}}}-10\right)~.
\label{final_pq_bound}
\end{eqnarray}
Of course, making $p_{\rm single}$ and $q_{\rm single}$ smaller can only make things better. Our conditions on $p_{\rm hook}$ and $q_{\rm hook}$ in eq.~(\ref{final_pq_bound}) are not smart enough to know this --- for $p_{\rm single}$ sufficiently small, we find that making it still smaller gives us a {\em more} stringent condition on $p_{\rm hook}$, and similarly for $q_{\rm hook}$. Clearly, this behavior is an artifact of our approximations. Thus, for a given $p_{\rm single}$ and $q_{\rm single}$, we are free to choose any smaller values of $p_{\rm single}$ and $q_{\rm single}$ in order to obtain  more liberal conditions on $p_{\rm hook}$ and $q_{\rm hook}$ from eq.~(\ref{final_pq_bound}). Our expression that bounds $p_{\rm hook}$ achieves its maximum for $p_{\rm single}= (3/40)^2$, and for fixed $p_{\rm single}$, our expression that bounds $q_{\rm hook}$ achieves its maximum for $q_{\rm single}=(1/20)^2$. We therefore conclude that
for recovery to succeed with a probability that approaches one as the block size increases, it suffices that
\begin{eqnarray}
&&p_{\rm single}< {9\over 1600}~, \quad q_{\rm single} <{1\over 400}  ~,\nonumber\\
&&p_{\rm hook} < {3\over 32}\cdot {9\over 1600}~,\quad q_{\rm hook} < {1\over 16}\cdot {9\over 1600}~.
\end{eqnarray}
Comparing to our expressions for $q_{\rm single}$, $p_{\rm single}$, and $p_{\rm hook}$, we see that, unless $q_{\rm single}$ is dominated by preparation or measurement errors, these conditions are all satisfied provided that
\begin{equation}
\label{gate_accuracy}
q_{\rm hook}=3p_{\rm CNOT} + 2p_s < 3.5 \times 10^{-4}~.
\end{equation}
If the probability of a CNOT error is negligible, then we obtain a lower bound on the critical error probability for storage errors, 
\begin{equation}
\left(p_s\right)_c > 1.7 \times 10^{-4}~.
\end{equation}
In view of the crudeness of our combinatorics, we believe that this estimate is rather conservative, if one accepts the assumptions of our computational model.

\section{Measurement and encoding}
\label{sec:enc_meas}

\subsection{Measurement}
\label{subsec:measurement}

At the conclusion of a quantum computation, we need to measure some qubits. If
the computation is being executed fault tolerantly, this means measuring an
encoded block.  How can we perform this measurement fault tolerantly?

Suppose we want to measure the logical operator $\bar Z$; that is, measure the
encoded block in the basis $\{\ket{\bar 0},\ket{\bar 1}\}$. If we are willing
to destroy the encoded block, we first measure $Z$ for each qubit in the block,
projecting each onto the basis $\{|0\rangle, |1\rangle\}$. Were there no errors
in the code block at the time of the measurement, and were all measurements of
the individual qubits performed flawlessly, then we could choose any
homologically nontrivial path on the lattice and evaluate the parity of the
outcomes for the links along that path.  Even parity indicates that the encoded
block is in the state $|\bar 0\rangle$, odd parity the state $|\bar 1\rangle$.

But the code block {\em will} contain some errors (not too many, we hope), and
some of the measurements of the individual qubits {\em will} be faulty. Since a
single bit flip along the path could alter the parity of the measurement
outcomes, we need to devise a fault-tolerant procedure for translating the
observed values of the individual qubits into a value of the encoded qubit.

One such procedure is to evaluate the parity $Z^{\otimes 4}$ of the measurement outcomes
at each plaquette of the lattice, determining the locations of all
plaquette defects. These defects can arise either because defects were already
present in the code block before the measurement, or they could be introduced
by the measurement itself.  It is useful and important to recognize that the
defects introduced by the measurement do not pose any grave difficulties.   An
isolated measurement error at a single link will produce two neighboring
defects on the pla\-quettes that contain that link.  Widely separated defects can
arise from the measurement only if there are many correlated measurement
errors.

Therefore we can apply a suitable classical algorithm to remove the defects --
for example by choosing a chain of minimal total length that is bounded by the
defect locations, which can be found in a polynomial-time classical
computation. Flipping the bits on this chain corrects the errors in the
measurement outcomes, so that we can then proceed to evaluate the parity along
a nontrivial cycle.  Assuming sufficiently small rates for the qubit and
measurement errors, the encoded qubit will be evaluated correctly, with a
probability of error that is exponentially small for large block size.

We can measure $\bar X$ by the same procedure, by measuring $X$ for each qubit,
and evaluating all site operators $X^{\otimes 4}$ from the outcomes. After removal of the site defects by flipping bits appropriately, $\bar X$ is the parity along a nontrivial cycle of the dual lattice.

To measure $\bar Z$ of a code block without destroying the encoded state, we
can prepare an ancilla block in the encoded state $|\bar 0\rangle$, and perform
a bitwise CNOT from the block to be measured into the ancilla.  Then we can
measure the ancilla by the destructive procedure just described.  A
nondestructive measurement of $\bar X$ is executed similarly.

\subsection{Encoding of known states}
\label{subsec:encoding_known}
At the beginning of a quantum computation, we need to prepare encoded qubits in eigenstates of the encoded operations, for example the state $|\bar 0\rangle$ of the planar
code, a $\bar Z=1$ eigenstate. If syndrome measurement were perfectly reliable, the state $|\bar0\rangle$ could be prepared quickly by the following method:  Start with the state
$|0\rangle^{\otimes n}$ where $n$ is the block size of the code.  This is the
simultaneous eigenstate with eigenvalue $1$ of all plaquette stabilizer
operators $Z_P=Z^{\otimes 4}$ and of the logical operator $\bar Z$, but not of the
site stabilizer operators $X_s=X^{\otimes 4}$.  Then measure all the site operators. Since the site operators commute with the plaquette operators and the logical operators, this measurement does not disturb their values. About half of the site measurements have outcome $X_s=1$ and about half have outcome $X_s=-1$; to obtain the state $|\bar 0\rangle$, we must remove all of the site defects (sites where $X_s=-1$). Thus we select an arbitrary 1-chain whose boundary consists of the positions of all site defects, and we apply
$Z$ to each link of this chain, thereby imposing $X_s=1$ at each
site. In carrying out this procedure, we might apply $\bar Z$ to the
code block by applying $Z$ to a homologically nontrivial path, but this has no
effect since the state is a $\bar Z=1$ eigenstate.

Unfortunately, syndrome measurement is not perfectly
reliable; therefore this procedure could generate long {\em open} chains of $Z$ errors
in the code block. To keep the open chains under control, we need to repeat the
measurement of both the $X$ and $Z$ syndromes  of order $L$ times (where $L$ is
the linear size of the lattice), and use our global recovery method. Then the initial configuration of the defects will be ``forgotten'' and the error chains in the code block will relax to the equilibrium configuration in which long open chains are highly unlikely. The probability of
an $\bar X$ error that causes a flip of the encoded state will be
exponentially small in $L$. We can prepare the encoded state with $\bar X=1$ by the dual
procedure, starting with the state $[{1\over \sqrt{2}}(|0\rangle +
|1\rangle)]^{\otimes n}$.
 
\subsection{Encoding of unknown states}
\label{subsec:enc_unknown}

Quantum error-correcting codes can protect {\em unknown} coherent quantum
states.  This feature is crucial in applications to quantum computation --- the
operator of a quantum computer need not ``monitor'' the encoded quantum state
to keep the computation on track.  But to operate a quantum computer, we don't
typically need to ${\em encode}$ unknown quantum states.  It is sufficient to
initialize the computer by encoding known states, and then execute a known
quantum circuit.

Still, a truly robust ``quantum memory'' should be able to receive an unknown
quantum state and store it indefinitely.  But given any nonzero rate of
decoherence, to store an unknown state for an indefinitely long time we need to
encode it using a code of indefinitely long block size.  How, then, can we
expect to encode the state before it decoheres?

The key is to encode the state quickly, providing some measure of protection,
while continuing to build up toward larger code blocks.  Concatenated codes
provide one means of achieving this. We can encode, perform error correction,
then encode again at the next level of concatenation.  If the error rates are
small enough, encoding can outpace the errors so that we can store the unknown
state in a large code block with reasonable fidelity.

The surface codes, too, allow us to build larger codes from smaller codes and
so to protect unknown states effectively. The key to enlarging the code block is that a code
corresponding to one triangulation of a surface can be transformed into a code corresponding to another triangulation.

For example, we can transform one surface code to another using local moves shown in
Fig.~\ref{fig:moves}:

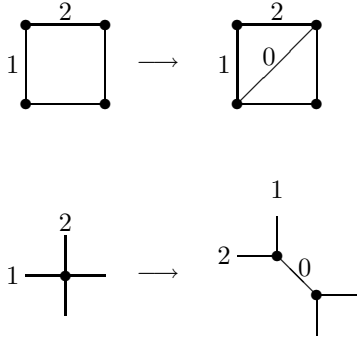
\begin{figure}[h]
\centering
\begin{picture}(140,140)

\put(5,105){\makebox(0,0){1}}
\put(25,125){\makebox(0,0){2}}

\put(10,90){\line(1,0){30}}
\put(40,90){\line(0,1){30}}
\put(40,120){\line(-1,0){30}}
\put(10,120){\line(0,-1){30}}

\put(10,90){\circle*{4}}
\put(10,120){\circle*{4}}
\put(40,90){\circle*{4}}
\put(40,120){\circle*{4}}

\put(60,105){\makebox(0,0){$\longrightarrow$}}

\put(85,105){\makebox(0,0){1}}
\put(105,125){\makebox(0,0){2}}
\put(102,108){\makebox(0,0){0}}

\put(90,90){\line(1,0){30}}
\put(120,90){\line(0,1){30}}
\put(120,120){\line(-1,0){30}}
\put(90,120){\line(0,-1){30}}
\put(90,90){\line(1,1){30}}

\put(90,90){\circle*{4}}
\put(90,120){\circle*{4}}
\put(120,90){\circle*{4}}
\put(120,120){\circle*{4}}

\put(25,10){\line(0,1){30}}
\put(10,25){\line(1,0){30}}
\put(25,25){\circle*{4}}
\put(5,25){\makebox(0,0){1}}
\put(25,45){\makebox(0,0){2}}

\put(60,25){\makebox(0,0){$\longrightarrow$}}

\put(90,32.5){\line(1,0){15}}
\put(105,32.5){\line(0,1){15}}
\put(105,32.5){\circle*{4}}

\put(105,32.5){\line(1,-1){15}}

\put(120,17.5){\circle*{4}}
\put(120,17.5){\line(1,0){15}}
\put(120,17.5){\line(0,-1){15}}

\put(85,32.5){\makebox(0,0){2}}
\put(105,57.5){\makebox(0,0){1}}
\put(115.5,28){\makebox(0,0){0}}

\end{picture}
\caption{Two basic moves that modify the triangulation of a surface by adding a
link: splitting a plaquette, and splitting a vertex.}
\label{fig:moves}
\end{figure}

\noindent Links can be added to (or removed from) the triangulation in either of two
ways --- one way adds a new plaquette, the other adds a new site.  Either way,
the new triangulation corresponds to a new code with an additional qubit in the code block and an
additional stabilizer generator.

When a new plaquette is added, the new code stabilizer is obtained
from the old one by adding the new plaquette operator
\begin{equation}
Z_1 Z_2 Z_0
\end{equation}
and by modifying the site operators with the replacements
\begin{equation}
\label{site_change}
X_1 \to X_1 X_0~, \quad X_2\to X_2 X_0~.
\end{equation}
When a new site is added, the stabilizer is modified similarly, but with
$X$'s and $Z$'s interchanged:
\begin{equation}
 X_1 X_2 X_0 ~,
\end{equation}
is a new stabilizer generator, and the existing plaquette operators are modified as
\begin{equation}
\label{plaq_change}
Z_1 \to Z_1 Z_0~, \quad Z_2\to Z_2 Z_0~.
\end{equation}

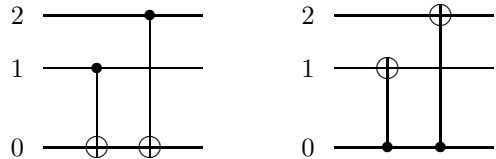
\begin{figure}[h]
\centering
\begin{picture}(180,80)

\put(0,10){\makebox(0,0){0}}
\put(0,40){\makebox(0,0){1}}
\put(0,60){\makebox(0,0){2}}

\put(10,10){\line(1,0){60}}
\put(10,40){\line(1,0){60}}
\put(10,60){\line(1,0){60}}

\put(30,40){\circle*{4}}
\put(30,40){\line(0,-1){34}}
\put(30,10){\circle{8}}

\put(50,60){\circle*{4}}
\put(50,60){\line(0,-1){54}}
\put(50,10){\circle{8}}

\put(110,10){\makebox(0,0){0}}
\put(110,40){\makebox(0,0){1}}
\put(110,60){\makebox(0,0){2}}

\put(120,10){\line(1,0){60}}
\put(120,40){\line(1,0){60}}
\put(120,60){\line(1,0){60}}

\put(140,10){\circle*{4}}
\put(140,10){\line(0,1){34}}
\put(140,40){\circle{8}}

\put(160,10){\circle*{4}}
\put(160,10){\line(0,1){54}}
\put(160,60){\circle{8}}

\end{picture}
\caption{Circuits that implement the two basic moves of Fig.~\ref{fig:moves}.
The circuit with qubit 0 as the target of the CNOT's adds a plaquette; the
circuit with qubit 0 as the control of the CNOT's adds a site.}
\label{fig:move_circuits}
\end{figure}

To add a plaquette or a site to a stabilizer code, we prepare the additional
qubit in a $Z_0=1$ or $X_0=1$ eigenstate, and then execute the circuit shown
in Fig.~\ref{fig:move_circuits}.
We recall that, acting by conjugation, a CNOT gate changes a tensor
product of Pauli operators acting on its control and target according to
\begin{equation}
IZ \leftrightarrow ZZ~,\quad XI \leftrightarrow XX~;
\end{equation}
that is, the CNOT transforms an
$IZ$ eigenstate to a $ZZ$ eigenstate and an $XI$ eigenstate to an $XX$ eigenstate, while leaving $ZI$ and $IX$ eigenstates
invariant. The circuit in Fig.~\ref{fig:move_circuits} with qubit 0 as target,
then, transforms the site operators as in eq.~(\ref{site_change}) while
also implementing
\begin{equation}
Z_0\to Z_1 Z_2 Z_0~.
\end{equation}
The initial $Z_0=1$ eigenstate is transformed into a state that satisfies the
plaquette parity checks of the new triangulation.   Similarly the circuit in
Fig.~\ref{fig:move_circuits} with qubit 0 as control implements
eq.~(\ref{plaq_change}) as well as
\begin{equation}
X_0\to X_1 X_2 X_0~;
\end{equation}
the circuit transforms the $X_0=1$ eigenstate into a state that satisfies the
new site parity checks.

Of course, these circuits are reversible; they can be used to extricate qubits
from a stabilizer code instead of adding them.

If planar codes are used, we can lay out the qubits in a planar array.
Starting with a small encoded planar block in the center, we can gradually add
new qubits to the boundary using the moves shown in
Fig.~\ref{fig:boundary_moves}:

\begin{figure}[h]
\centering
\begin{picture}(240,120)

\put(25,5){\makebox(0,0){2}}
\put(35,25){\makebox(0,0){1}}

\put(10,40){\line(1,0){90}}
\put(10,40){\line(0,-1){30}}
\put(10,40){\line(0,1){10}}
\put(40,40){\line(0,-1){28}}
\put(40,40){\line(0,1){10}}
\put(70,40){\line(0,-1){30}}
\put(70,40){\line(0,1){10}}
\put(100,40){\line(0,-1){30}}
\put(100,40){\line(0,1){10}}
\put(10,10){\line(1,0){28}}
\put(10,40){\circle*{4}}
\put(40,40){\circle*{4}}
\put(70,40){\circle*{4}}
\put(100,40){\circle*{4}}
\put(10,10){\circle*{4}}
\put(40,10){\circle{4}}

\put(120,25){\makebox(0,0){$\longrightarrow$}}

\put(140,40){\line(1,0){90}}
\put(140,40){\line(0,-1){30}}
\put(140,40){\line(0,1){10}}
\put(170,40){\line(0,-1){30}}
\put(170,40){\line(0,1){10}}
\put(200,40){\line(0,-1){28}}
\put(200,40){\line(0,1){10}}
\put(140,10){\line(1,0){30}}
\put(170,10){\line(1,0){28}}
\put(230,10){\line(0,1){30}}
\put(230,40){\line(0,1){10}}
\put(140,10){\circle*{4}}
\put(170,10){\circle*{4}}
\put(200,40){\circle*{4}}
\put(200,10){\circle{4}}
\put(230,40){\circle*{4}}
\put(140,40){\circle*{4}}
\put(170,40){\circle*{4}}

\put(155,5){\makebox(0,0){2}}
\put(165,25){\makebox(0,0){1}}
\put(185,5){\makebox(0,0){0}}

\put(55,105){\makebox(0,0){1}}
\put(35,85){\makebox(0,0){2}}

\put(10,100){\line(1,0){90}}
\put(10,100){\line(0,-1){30}}
\put(10,100){\line(0,1){10}}
\put(40,100){\line(0,-1){30}}
\put(40,100){\line(0,1){10}}
\put(70,100){\line(0,1){10}}
\put(100,100){\line(0,1){10}}
\put(10,100){\circle*{4}}
\put(40,100){\circle*{4}}
\put(70,100){\circle*{4}}
\put(100,100){\circle*{4}}
\put(55,85){\circle{4}}

\put(120,85){\makebox(0,0){$\longrightarrow$}}

\put(140,100){\line(1,0){90}}
\put(140,100){\line(0,-1){30}}
\put(140,100){\line(0,1){10}}
\put(170,100){\line(0,-1){30}}
\put(170,100){\line(0,1){10}}
\put(200,100){\line(0,-1){30}}
\put(200,100){\line(0,1){10}}
\put(230,100){\line(0,1){10}}
\put(170,100){\line(1,0){30}}
\put(140,100){\circle*{4}}
\put(170,100){\circle*{4}}
\put(200,100){\circle*{4}}
\put(230,100){\circle*{4}}
\put(215,85){\circle{4}}

\put(185,105){\makebox(0,0){1}}
\put(165,85){\makebox(0,0){2}}
\put(195,85){\makebox(0,0){0}}

\end{picture}
\caption{The same circuits as in Fig.~\ref{fig:move_circuits} can also be used
to build up a planar code by adding a link at the boundary. Sites or plaquettes
marked by open circles do not correspond to stabilizer operators.}
\label{fig:boundary_moves}
\end{figure}
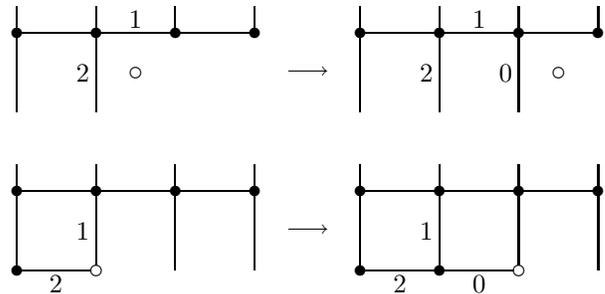

\noindent These moves add a new three-qubit plaquette or site operator, and
can also be implemented by the circuits of Fig.~(\ref{fig:move_circuits}).

\begin{figure}
\begin{center}
\leavevmode
\epsfxsize=4in
\epsfbox{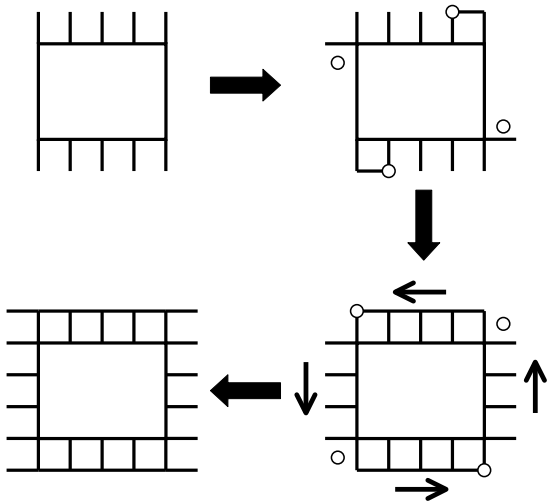}
\end{center}
\caption{Building a distance-$(L+1)$ planar code by adding qubits to a
distance-$L$ planar code. (Here, $L=5$.) In the first
step,  new two-qubit stabilizer operators are added in the corners with
single CNOT's; in subsequent steps, three-qubit stabilizer operators are added
with double CNOT's. The last step promotes the corner operators to three-qubit
operators.}
\label{fig:build_up}
\end{figure}

A procedure that transforms a distance-$L$ planar code to a distance-($L+1$)
code is shown in Fig.~\ref{fig:build_up}.
By adding a new row of plaquette operators, we transform what was formerly a smooth edge into a rough edge, and by adding a new row of site operators we transform a rough edge to a smooth edge. We start the row of plaquettes by adding a two-qubit plaquette operator to the corner via the transformations 
\begin{equation}
Z_0 \to Z_1 Z_0~, \quad X_1\to X_1 X_0 ~, 
\end{equation}
which can be implemented by a single CNOT;
similarly, we start a row of sites by adding a two-qubit site operator with 
\begin{equation}
X_0 \to X_1 X_0~, \quad Z_1\to Z_1 Z_0 ~. 
\end{equation}
Then  a new row of boundary stabilizer operators can be ``zipped'' into place.

As is typical of encoding circuits, this procedure can propagate errors badly;
a single faulty CNOT can produce a long row of qubit errors (a widely separated
pair of defects) along the edge of the block.  To ensure fault tolerance, we
must measure the boundary stabilizer operators frequently during the procedure.
 Examining the syndrome record, we can periodically identify the persistent
errors and remove them before proceeding to add further qubits.

\section{Fault-tolerant quantum computation}
\label{sec:ftgates}
We will now consider how information protected by planar surface codes can be processed fault-tolerantly. Our objective is to show that a universal set of fault-tolerant encoded quantum gates can be realized using only local quantum gates among the fundamental qubits and with only polynomial overhead. We will describe one gate set with this property \cite{kitaev1,kitaev_threshold}. This construction suffices to show that there is an accuracy threshold for quantum computation using surface codes: each gate in our set can be implemented acting on encoded states with arbitrarily good fidelity, in the limit of a large code block. We have not analyzed the numerical value of this computation threshold in detail. Better implementations of fault-tolerant quantum computation can probably be found, requiring less overhead and yielding a better threshold.

We choose the basis introduced by Shor \cite{shor_ft}, consisting of four gates. Three of these generate the ``symplectic'' or ``normalizer'' group, the finite subgroup of the unitary group that, acting by conjugation, takes tensor products of Pauli operators to tensor products of Pauli operators. Of these three, two are single-qubit gates: the Hadamard gate
\begin{equation}
H={1\over \sqrt{2}}\pmatrix{1&1\cr 1&-1}~,
\end{equation}
which acts by conjugation on Pauli operators according to
\begin{equation}
H: X\leftrightarrow Z~,
\end{equation}
and the phase gate
\begin{equation}
P\equiv \Lambda(i)= \pmatrix{1&0\cr 0&i}~,
\end{equation}
which acts by conjugation on Pauli operators according to 
\begin{equation}
P: X\to Y~,\quad Z\to Z~.
\end{equation}
The third generator of the normalizer group is the two-qubit ${\rm CNOT}=\Lambda(X)$ gates, which acts by conjugation on Pauli operators according to
\begin{eqnarray}
\label{CNOT_conj}
{\rm CNOT}: &XI \to XX~,\quad &IX \to IX~,\nonumber\\
&ZI\to ZI~,\quad &IZ\to ZZ~.
\end{eqnarray}

Quantum computation in the normalizer group is no more powerful than classical computation \cite{gott_heisenberg}. To realize the full power of quantum computing we need to complete the basis with a gate outside the normalizer group. This gate can be chosen to be the three-qubit Toffoli gate $T\equiv \Lambda^2(X)$, which acts on the standard three-qubit orthonormal basis $\{|a,b,c\rangle\}$ as
\begin{equation}
T:|a,b,c\rangle \to |a,b,c\oplus ab\rangle~.
\end{equation}

\subsection{Normalizer gates for surface codes}

\subsubsection{CNOT gate}
Implementing normalizer computation on planar codes is relatively simple. First of all, a planar surface code is a Calderbank-Shor-Steane \cite{cal_shor,steane_css} (CSS) code, and as for any CSS code with a single encoded qubit, an encoded CNOT can be performed {\em transversally} --- in other words, if simultaneous CNOT's are executed from each qubit in one block to the corresponding qubit in the other block, the effect is to execute the encoded CNOT \cite{gott_ft}. To see this, we first need to verify that the transversal CNOT preserves the code space, {\it i.e.}, that its action by conjugation preserves the code's stabilizer. This follows immediately from eq.~(\ref{CNOT_conj}), since each stabilizer generator is either a tensor product of $X$'s or a tensor product of $Z$'s. Next we need to check that ${\rm CNOT}^{\otimes n}$ acts on the encoded operations $\bar X$ and $\bar Z$ as in eq.~(\ref{CNOT_conj}), which also follows immediately since $\bar Z$ is a tensor product of $Z$'s and $\bar X$ is a tensor product of $X$'s.

\subsubsection{Hadamard gate}

\begin{figure}
\begin{center}
\leavevmode
\epsfxsize=3in
\epsfbox{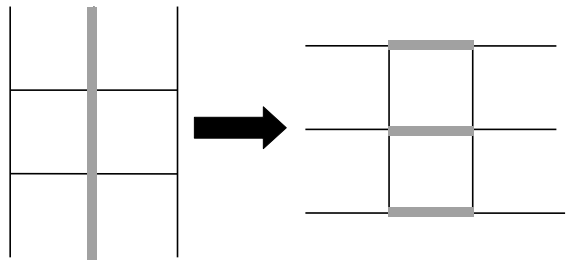}
\end{center}
\caption{Action of the bitwise Hadamard gate on the planar code. If Hadamard gates are applied simultaneously to all the qubits in the block, an ``NS code'' with rough edges at the north and south is transformed to an ``EW code'' with rough edges at the east and west; the encoded operation $\bar Z_{\rm NS}$ of the NS code is transformed to $\bar X_{\rm EW}$ of the EW code, and $\bar X_{\rm NS}$ is transformed to $\bar Z_{\rm EW}$.}
\label{fig:hadamard}
\end{figure}

What about the Hadamard gate? In fact, applying the bitwise operation $H^{\otimes n}$ does not preserve the code space; rather it maps the code space of one planar code to that of another, different, planar code. If the stabilizer generators of the initial code are site operators $X_s$ and plaquette operators $Z_P$, then the action of the bitwise Hadamard is
\begin{equation}
H^{\otimes n}: X_s \to Z_s~,\quad Z_P \to X_P~
\end{equation}
Compared to the initial code, the stabilizer of the new code has sites and plaquettes interchanged. We may reinterpret the new code as a code with $X_s$ and $Z_P$ check operators, but defined on a lattice dual to the lattice of the original code.
If the original lattice has its ``rough'' edges at the north and south, then the new lattice has its rough edges at the east and west. We will refer to the two codes as the ``north-south'' (NS) code and the ``east-west'' (EW) code. As indicated in Fig.~\ref{fig:hadamard}, the action of $H^{\otimes n}$ on the encoded operations $\bar X$ and $\bar Z$ of the NS code is
\begin{equation}
H^{\otimes n}: \bar X_{\rm NS}\to \bar Z_{\rm EW}~,\quad  \bar Z_{\rm NS}\to \bar X_{\rm EW}~.
\end{equation}
If we rigidly rotate the lattice by $90^\circ$, the EW code is transformed back to the NS code. Hence, the overall effect of a bitwise Hadamard and a $90^\circ$ rotation is an encoded Hadamard $\bar H$.

Of course, a physical rotation of the lattice might be inconvenient in practice! Instead, we will  suppose that ``peripheral'' qubits are available at the edge of the code block, and that we have the option of incorporating these qubits into the block or ejecting them from the block using the  method described in Sec.~\ref{subsec:enc_unknown}. After applying the bitwise Hadamard, transforming the $L\times L$ NS code to the EW code, we add $L-1$ plaquettes to the northern edge and $L-1$ sites to the western edge, while removing $L-1$ plaquettes on the east and $L-1$ sites on the south. This procedure transforms the block back to the NS code, but with the qubits shifted by half a lattice spacing to the north and west --- we'll call this shifted code the ${\rm NS}'$ code. Furthermore, this modification of the boundary transforms the logical operations $\bar Z_{\rm EW}$ and $\bar X_{\rm EW}$ of the EW code to the operations $\bar Z_{{\rm NS}'}$ and $\bar X_{{\rm NS}'}$ of the ${\rm NS}'$ code. The overall effect, then, of the bitwise Hadamard followed by the boundary modification is the operation
\begin{equation}
\bar X_{\rm NS}\to \bar Z_{{\rm NS}'}~,\quad  \bar Z_{\rm NS}\to \bar X_{{\rm NS}'}~.
\end{equation}
In principle, we could complete the encoded Hadamard gate by physically shifting the qubits half a lattice spacing to the south and east, transforming the ${\rm NS}'$ code back to the NS code. One way to execute this shift might be to swap the qubits of the ${\rm NS}'$ with qubits located at the corresponding sites of the NS lattice. If we prefer to avoid the additional quantum processing required by the swaps, then what we can do instead is associate a classical flag bit with each code block, recording whether the number of Hadamard gates that have been applied in our circuit to that logical qubit is even or odd, and hence whether the logical qubit is encoded in the NS code or the ${\rm NS}'$ code. This classical bit is consulted whenever the circuit calls for a Hadamard or CNOT acting on the block. 
If we perform a Hadamard on a qubit that is initially encoded with the ${\rm NS}'$ code, we add qubits on the south and east while removing them from the north and west, returning to the NS code. The CNOT gates are performed transversally between blocks that are both in the NS code or both in the ${\rm NS}'$ code; that is, each qubit in one layer interacts with the corresponding qubit directly below it in the next layer. But if one block is in the NS code and the other is in the ${\rm NS}'$ code, then each qubit in one layer interacts with the qubit in the next layer that is half a lattice spacing to north and west. Note that the modification of the boundary requires a number of computation steps that is linear in $L$. 

\subsubsection{Phase gate}
For implementation of the phase gate $P$, note that if we can execute CNOT and $H$ then we can also construct the ``controlled-$(iY)$'' gate 
\begin{equation}
\Lambda(iY)= \Lambda(ZX)= (IH)\cdot\Lambda(X)\cdot(IH)\cdot\Lambda(X)~.
\end{equation}
Hence it suffices to be able to prepare an eigenstate $|+\rangle$ or $|-\rangle$ of $Y$,
\begin{equation}
Y|\pm\rangle = \pm |\pm\rangle~;
\end{equation}
if we prepare an ancilla in the state $|+\rangle$, and apply a CNOT with the data as its control and the ancilla as its target, the effect on the data is the same as $\Lambda(i)=P$. If the ancilla is the state $|-\rangle$, then we apply $\Lambda(-i)=P^{-1}$ to the data instead.

Now, it is not obvious how to prepare a large toric block in an eigenstate of the encoded $ Y$ with good fidelity. Fortunately, we can nevertheless use a CNOT and an ancilla to implement $P$, thanks to a trick that works because $P$ is the only gate in our set that is not real. Consider a circuit that applies the unitary transformation $U$ to the data if the ancilla has actually been prepared in the state $|+\rangle$. Then if $|+\rangle$ were replaced by $|-\rangle$, this same circuit would apply the complex conjugate unitary $U^*$, since each $P$ in the circuit would be replaced by $P^*$.

Instead of  a $Y$ eigenstate, suppose we prepare the ancilla in any encoded state we please, for example $|\bar 0\rangle$. And then we use this same ancilla block, and a CNOT, every time a $P$ is to be executed. The state of the ancilla can be expressed as a linear combination $a|+\rangle + b |-\rangle$ of the $Y$ eigenstates, and our circuit, acting on the initial state $|\psi\rangle$ of the data, yields 
\begin{equation}
a|+\rangle \otimes U|\psi\rangle + b|-\rangle\otimes U^*|\psi\rangle~.
\end{equation}
Now, at the very end of a quantum computation, we will need to make a measurement to read out the final result. Let $A$ denote the observable that we measure. The expectation value of $A$ will be 
\begin{equation}
\langle A\rangle = |a|^2 \langle\psi |U^\dagger A U|\psi\rangle +|b|^2 \langle \psi| U^\dagger A^T U|\psi\rangle~,
\end{equation}
where $A^T$ denotes the transpose of $A$. Without losing any computational power, we may assume that the observable $A$ is real ($A=A^T$) --- for example it could be ${1\over 2}(I-Z)$ acting on one of our encoded blocks. Then we get the same answer for the expectation value of $A$ as if the ancilla had been prepared as $|+\rangle$ (or $|-\rangle$); hence our fault-tolerant procedure successfully simulates the desired quantum circuit.

Since there is just one ancilla block that must be used each time the $P$ gate is executed, this block has to be swapped into the position where it is needed, a slowdown that is linear in the width of the quantum circuit that is being simulated.

Thus we have described a way to perform fault-tolerant normalizer computation for planar surface codes. We envision, then, a quantum computer consisting of a stack of planar sheets, with a logical qubit residing in each sheet. Each logical sheet has associated with it an adjacent sheet of ancilla qubits that are used to measure the check operators of the surface code; after each measurement, these ancilla qubits are refreshed in place and then reused. The quantum information in one sheet can be swapped with that in the neighboring sheet through the action of local gates. To perform a logical CNOT between two different logical qubits in the stack, we  first use swap gates to pass the qubits through the intervening sheets of logical and ancilla qubits and bring them into contact, then execute the transversal CNOT between the two layers, and then use swap gates to return the logical qubits to their original positions. By inserting a round of error correction after each swap or logical operation, we can execute a normalizer circuit reliably.

\subsection{State purification and universal quantum computation}

Now we need to consider how to complete our universal gate set by adding the Toffoli gate. As Shor observed \cite{shor_ft}, implementation of the gate can be reduced to the problem of preparing a particular three-qubit state, which may be chosen to be
\begin{equation}
|\psi\rangle_{\rm anc}=2^{-3/2}\sum_{a,b,c\in\{0,1\}} (-1)^{abc}|a\rangle_1|b\rangle_2|c\rangle_3~;
\end{equation}
this state is the simultaneous eigenstate of three commuting symplectic operators: 
$\Lambda(Z)_{1,2}X_3$ and its two cyclic permutations, where $\Lambda(Z)$ is the two-qubit conditional phase gate
\begin{equation}
\Lambda(Z): |a,b\rangle \to (-1)^{ab}|a,b\rangle
\end{equation}
Shor's method for constructing this state involved the preparation and measurement of an unprotected $n$-qubit cat state, where $n$ is the block size of the code. But this method cannot be used for a toric code on a large lattice, because the cat state is too highly vulnerable to error.

Fortunately, there is an alternative procedure for constructing the needed encoded state with high fidelity --- {\em state purification}. Suppose that we have a supply of noisy copies of the state $|\psi\rangle_{\rm anc}$. We can carry out a purification protocol to distill from our initial supply of noisy states a smaller number of states with much better fidelity \cite{kit_purify,dennis}. In this protocol, normalizer gates are applied to a pair of noisy copies, and then one member of the pair is measured. Based on the outcome of the measurement, the other state is either kept or discarded. If the initial ensemble of states approximates the $|\psi\rangle_{\rm anc}$ with adequate fidelity, then as purification proceeds, the fidelity of the remaining ensemble converges rapidly toward one.

For this procedure to work, it is important that our initial states are not {\em too} noisy --- there is a purification threshold. Therefore, to apply the purification method to toric codes, we will need to build up the size of the toric block gradually, as in the procedure for encoding unknown states described in Sec.~\ref{subsec:enc_unknown}. We start out by encoding $|\psi\rangle_{\rm anc}$ on a small planar sheet of qubits, with a fidelity below the purification threshold. Then we purify for a while to improve the fidelity, and build on the lattice to increase the size of the code block. By building and purifying as many times as necessary, we can construct a copy of the ancilla state that can be used to execute the Toffoli gate with high fidelity.

The time needed to build up the encoded blocks is quadratic in $L$, and the number of rounds of purification needed is linear in $L$, if we wish to reach a fidelity that is exponentially small in $L$. Thus the overhead incurred in our implementation of the Toffoli gate is polynomial in the block size.

We have now assembled all the elements of a fault-tolerant universal quantum computer based on planar surface codes. The computer is a stack of logical qubits, and it contains  ``software factories'' where the ancilla states needed for execution of the Toffoli gate are prepared. Once prepared, these states can be transported through swapping to the position in the stack where the Toffoli gate is to be performed.

\section{A local algorithm in four dimensions}
\label{sec:four_dim}

In our recovery procedure, we have distinguished between quantum and classical
computation.  Measurements are performed to collect syndrome information about
errors that have accumulated in the code block, and then a fast and reliable
classical computer processes the measured data to infer what recovery step is
likely to remove most of the errors.  Our procedures are fault tolerant because the
quantum computation needed to measure the syndrome is
highly local.  But the classical computation not so local --- our algorithm for constructing the chain of minimal weight requires as input the syndrome history of the entire code block. 

It would be preferable to replace this procedure by one in which measurements and classical processing are eliminated, and all of the processing is local quantum processing. Can we devise a stable quantum memory based on topological coding such that rapid measurements of the syndrome are not necessary?

Heuristically, errors create pairs of defects in the code block, and trouble may arise if these defects diffuse apart and annihilate other defects, eventually generating homologically nontrivial defect world lines. In principle, we could protect the encoded quantum information effectively if there is a strong attractive interaction between defects that prevents them from wandering apart. A recovery procedure that simulates such interactions was discussed in Ref.~\cite{dennis}. For that procedure, an accuracy threshold can be established, but only if the interactions have arbitrarily long range, in which case the order-disorder transition in the code block is analogous to the Kosterlitz-Thouless transition in a two-dimensional Coulomb gas. But to simulate these infinite-range interactions, nonlocal processing is still required.

A similar problem confronts the proposal \cite{kitaev2,ogburn,freedman} to encode quantum information in a configuration of widely separated nonabelian anyons. Errors create anyons in pairs, and the encoded information is endangered if these ``thermal anyons'' diffuse among the anyons that encode the protected quantum state. In principle, a long-range attractive interaction among anyons might control the diffusion, but this interaction might also interfere with the exchanges of anyons needed to process the encoded state. In any case, a simulation of the long-range dynamics involves nonlocal processing.

We will now describe a procedure for recovery that, at least mathematically,
requires no such nonlocal processing of quantum or classical information. With
this procedure, based on ``locally available'' quantum information, we can
infer a recovery step that is more likely to remove errors than add new ones.
Because the procedure is local we can dispense
with measurement without degrading its performance very much --- measurements followed by quantum gates conditioned on measurement outcomes can be replaced my unitary transformations acting on the data qubits and on nearby ancilla qubits. But since we
will still need a reservoir where we can dispose the entropy introduced by
random errors, we will continue to assume as usual that the ancilla qubits can be regularly refreshed as needed. 

Unfortunately, while our procedure is local in the mathematical sense that
recovery operations are conditioned on the state of a small number of
``nearby'' qubits, we do not know how to make it {\em physically} local in a
space of fewer than four dimensions.

\subsection{Repetition code in two dimensions}
The principle underlying our local recovery procedure can be understood if we
first consider the simpler case of a repetition code.  We can imagine
that the code block is a periodically identified one-dimensional lattice of
binary spins, with two codewords corresponding to the configurations with all
spins up or all spins down.  To diagnose errors, we can perform a local
syndrome measurement by detecting whether each pair of neighboring spins is aligned or
anti-aligned, thus finding the locations of defects where the spin orientation
flips.

To recover we need to bring these defects together in pairs to annihilate.  One
way to do this is to track the history of the defects for a while, assembling a
record $S$ of the measured syndrome, and then find a minimum-weight
chain $E'$ with the same boundary, in order to reconstruct hypothetical world lines
of the defects.  But in that case the processing required to construct $E'$ is nonlocal.

\begin{figure}
\begin{center}
\leavevmode
\epsfxsize=3in
\epsfbox{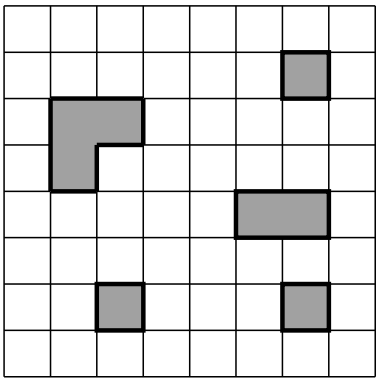}
\end{center}
\caption{Droplets of flipped qubits in the two-dimensional quantum repetition code. Qubits reside on plaquettes, and the qubits that have been flipped are shaded. Thick links are locations of ``defects'' where the error syndrome is nontrivial because neighboring qubits are anti-aligned. The defects form closed loops that enclose the droplets.} 
\label{fig_droplets}
\end{figure}

The way to attain a local recovery procedure is to increase the dimensionality
of the lattice. In two dimensions, errors will generate droplets of flipped
spins (as in Fig.~\ref{fig_droplets}), and the local syndrome measurement will detect the boundary of the
droplet.  Thus the defects now form one-dimensional closed loops, and our
recovery step should be designed to reduce the total length of such defects.
Local dynamical rules can easily be devised that are more likely to shrink a loop than
stretch it, just as it is possible to endow strings with local dynamics (tension and
dissipation) that allow the strings to relax.  Thus in equilibrium, very long
loops will be quite rare. If the error rate is small enough, then the droplets
of flipped spins will typically remain small, and the encoded information will
be well protected.

That the two-dimensional version of the repetition code is more robust than the one-dimensional version illustrates a central principle of statistical mechanics --- that order is more resistant to fluctuations in higher dimensions. The code block is described by an Ising spin model, and while the one-dimensional Ising model is disordered at any nonzero temperature, the two-dimensional Ising model remains ordered up to a nonvanishing critical temperature. From the perspective of coding theory, the advantage of the two-dimensional version is that the syndrome is highly redundant. If we check each pair of nearest-neighbor spins to see if they are aligned or anti-aligned, we are collecting more information than is really needed to diagnose all the errors in the block. Hence there is a constraint that must be satisfied by a valid syndrome, namely that the boundary of a droplet can never end; therefore errors in the syndrome can be detected. Of course, physically, the stability of the ordered state of the Ising model in more than one dimension is the reason that magnetic memories are robust in Nature.

\subsection{Toric code in four dimensions}
The defects detected by the measurement of the stabilizer operators of a two-dimensional toric
code are also pointlike objects, and error recovery is achieved by bringing the
defects together to annihilate. We can promote the annihilation by introducing
an effective long-range interaction between defects, but a more local alternative procedure
is to increase the dimensionality of the lattice.

So consider a {\em four-dimensional} toric code.  Qubits are associated with each
plaquette.  With each link is associated the six-qubit stabilizer operator
$X_\ell=X^{\otimes 6}$ acting on the six plaquettes that contain the link, and with each cube
is associated the six-qubit stabilizer operator $Z_C=Z^{\otimes 6}$ acting on the six
plaquettes contained in the cube.  Thus the four-dimensional code maintains the
duality between phase and flip errors that we saw in two dimensions.  The
encoded $\bar Z$ or $\bar X$ operation is constructed from $Z$'s or $X$'s
acting on a homologically nontrivial surface of the lattice or dual lattice
respectively.   $Z$ errors on a connected open surface generate a closed loop
of defects on the boundary of the surface, and $X$ errors on a connected open
surface of the dual lattice generate defects on a set of cubes that form a
closed loop on the dual lattice. As in the two-dimensional case, there is a ``hyperplanar'' version of the code that can be defined on a four-dimensional region with a boundary.

Now we want to devise a recovery procedure that will encourage the defect loops
to shrink and disappear.  Assuming that syndrome measurements are employed, a
possible procedure for controlling phase errors can be described as follows:
First, the stabilizer operator $X_\ell$ is measured at each link, and a record
is stored of the outcome.  We say that each link with $X_\ell=-1$ is occupied
by a string, and each link with $X_\ell=1$ is unoccupied.  We choose a set of
nonoverlapping plaquettes (with no link shared by two plaquettes in the set),
and based on the syndrome for the links of that plaquette, decide whether or
not to flip the plaquette (by applying a $Z$).  If three or four of the
plaquette's links are occupied by string, we always flip the plaquette.  If
zero or one link is occupied, we never flip it.  And if two links are occupied,
we flip the plaquette with probability $1/2$.  Then in the next time step, we
again measure the syndrome, and decide whether to flip another nonoverlapping
set of plaquettes.  And so on.

Naturally, we also measure the bit-flip syndrome --- $Z_C$ on every cube ---
in each time step. The procedure for correcting the bit-flip errors is
identical, with the lattice replaced by the dual lattice, and $X$ replaced by
$Z$.

Of course the measurement is not essential.  A simple reversible computation
can imprint the number of string bits bounding a plaquette on ancilla qubits,
and subsequent unitary gates controlled by the ancilla can ``decide'' whether
to flip the plaquette. Note that a CNOT that is applied with probability $1/2$,
needed in the event that the plaquette has two string bits on its boundary, can
be realized by a Toffoli gate, where one of the control qubits is a member of a
Bell pair so that the control takes the value $1$ with probability $1/2$.

This recovery procedure has the property that, if it is perfectly executed and
no further errors occur during its execution, it will never increase the total
length of string on the lattice, but it will sometimes reduce the length.
Indeed, if it is applied repeatedly while no further errors occur, it will
eventually eliminate every string.  We have chosen to make the procedure
nondeterministic in the case where there are two string bits on a plaquette,
because otherwise the procedure would have closed orbits --- some string
configurations would oscillate indefinitely rather than continuing to shrink
and annihilate.  With the nondeterministic procedure, a steady state can be
attained only when all the strings have disappeared.

Actually, following the ideas of Toom \cite{toom}, it is possible to devise {\em anisotropic deterministic} procedures that also are guaranteed to remove all strings. These procedures, in fact, remove the strings more efficiently than our nondeterministic one, but are a little more difficult to analyze.

Of course, the recovery procedure will not really be executed flawlessly, and further errors
will continue to accumulate. Still, as error recovery is performed many times,
an equilibrium will eventually be attained in which string length is being
removed by recovery as often as it is being created by new errors.  If the
error rates are small enough, the equilibrium population of long string loops
will be highly suppressed, so that the encoded quantum information will be well
protected.

Eventually, say at the conclusion of a computation, we will want to measure
encoded qubits.  This measurement procedure does have a nonlocal component (as
the encoded information is topological), and for this purpose only we will
assume that a reliable classical computer is available to help with the
interpretation of the measured data. To measure the logical operator $\bar Z$,
say, we first measure every qubit in the code block.  Then we apply a classical
parity check, evaluating $Z_C$ for each cube of the lattice, thereby generating a
configuration of closed defect loops on the dual lattice.  To complete the
measurement, we first eliminate the defects by applying flips to a set of
plaquettes bounded by each loop. Then we can evaluate the product of $Z$'s
associated with a homologically nontrivial surface to find the value of $\bar
Z$.

Of course, when we eliminate the defects, we need to make sure that we choose
correctly among the homologically inequivalent surfaces bounded by the observed
strings.  One way to do so,  which is unlikely to fail when qubit and measurement
error probabilities are small, is to invoke the relaxation algorithm formulated
above to the classical measurement outcome.  Since our classical computer is reliable,
the algorithm eventually removes all strings, and then the value of $\bar Z$
can be determined.

\subsection{Accuracy threshold}
To evaluate the efficacy of the local recovery method, we need to find
the equilibrium distribution of defects. This equilibrium configuration
is not so easily characterized, but it will suffice to analyze a less effective
algorithm that does attain a simple steady state --- the heat bath algorithm.
To formulate the heat bath algorithm, suppose that strings carry an energy per
lattice unit length that we may normalize to one, and suppose that each
plaquette is in contact with a thermal reservoir at inverse temperature
$\beta$. In each time step, plaquettes are updated, with the change in the
string length bounding a plaquette governed by the Boltzmann probability
distribution.  Thus survival or creation of a length-4 loop is suppressed by
the factor
\begin{equation}
\label{boltz_04}
{{\rm Prob}(0\to4)\over {\rm Prob}(0\to 0)}= {{\rm Prob}(4\to 4)\over {\rm
Prob}(4\to 0)}= e^{-4\beta}~.
\end{equation}
Similarly, the probability of a plaquette flip when the length of bounding
string is 3 or 1 satisfies
\begin{equation}
\label{boltz_13}
{{\rm Prob}(1\to3)\over {\rm Prob}(1\to 1)}= {{\rm Prob}(3\to 3)\over {\rm
Prob}(3\to 1)}= e^{-2\beta}~.
\end{equation}
In the case of a plaquette with two occupied links, we again perform the flip
with probability 1/2. As before, this ensures ergodicity --- any initial
configuration has some nonvanishing probability of reaching any final
configuration.

Damage to encoded information arises from string ``world sheets'' that are
homologically nontrivial.  At low temperature, string loops are dilute and
failure is unlikely, but at a critical temperature the strings ``condense,''
and the encoded data are no longer well protected.  The critical temperature is
determined by a balance between Boltzmann factor $e^{-\beta \ell}$ suppressing a
string of length $\ell$ and the string entropy. The abundance of self-avoiding
closed loops of length $\ell$ behaves like \cite{madras}
\begin{equation}
n_{\rm SAW}^{(4)}(\ell)\sim P_4(\ell)(\mu_4)^\ell~,\quad \mu_4 \approx 6.77~,
\end{equation}
in $d=4$ dimensions, where $P_4(\ell)$ is a polynomial. Thus, large loops are rare
when the sum
\begin{equation}
\sum_\ell n_{\rm SAW}^{(4)}(\ell) e^{-\beta \ell} \sim \sum_\ell P_4(\ell) \left(\mu_4
e^{-\beta}\right)^\ell
\end{equation}
converges, and the system is surely ordered for $e^{-\beta}<\mu_4^{-1}$. Thus the critical inverse temperature $\beta_c$ satisfies
\begin{equation}
e^{-\beta_c}\ge (\mu_4)^{-1}~.
\label{critical_beta}
\end{equation}

Now, our local recovery procedure will not be precisely a heat bath algorithm.
But like the heat bath algorithm it is more likely to destroy string than
create it, and we can bound its performance by assigning to it an effective
temperature. For example, if no new errors arise and the algorithm is perfectly
executed, it will with probability one remove a length-4 string loop bounding a
plaquette.  In practice, though, the plaquette may not flip when the recovery
computation is performed, either because of a fault during its execution, or
because other neighboring plaquettes have flipped in the meantime.  Let us
denote by $q_4$ the probability that a plaquette, occupied by four string bits
at the end of the last recovery step, does not in fact flip during the current
step.  Similarly, let $q_3$ denote the probability that a plaquette with three string
bits fails to flip, and let $q_1$, $q_0$ denote the probabilities that plaquettes
containing one or zero string bits {\em do} flip. These quantities can all be
calculated, given the quantum circuit for recovery and a stochastic error
model.

Now we can find a positive quantity $q$ such that
\begin{eqnarray}
& & q_0, q_4\le q/(1+q)~,\nonumber\\
& & q_1, q_3\le \sqrt{q}/(1+\sqrt{q})~.
\end{eqnarray}
Comparing to eqs.~(\ref{boltz_04},\ref{boltz_13}), we see that our recovery algorithm is at least as effective as a heat bath algorithm
with the equivalent temperature
\begin{equation}
e^{-4\beta}=q~;
\end{equation}
in equilibrium strings of length $\ell$ are therefore suppressed by a factor no larger
than $e^{-\beta \ell}=q^{\ell/4}$. From our estimate of the critical temperature eq.~(\ref{critical_beta}), we
then obtain a lower bound on the  critical value of $q$:
\begin{equation}
q_c \ge (\mu_4)^{-4}\approx 4.8\times 10^{-4}~.
\end{equation}
This quantum system with local interactions has an accuracy threshold.

A local procedure that controls the errors in a quantum memory is welcome, but it is disheartening that four spatial dimensions are required. Of course, the four-dimensional code block can be projected to $d<4$ dimensions, but then interactions among four-dimensional neighbors become interactions between qubits that are distance $L^{(4-d)/d}$ apart, where $L$ is the linear size of the lattice. In a three-dimensional version of the toric code, we can place qubits on plaquettes, and associate check operators with links and cubes. Thus, phase error defects are strings and bit-flip error defects are point particles, or vice versa. Then we can recover locally (without measurement or classical computation) from either the phase errors or the bit-flip errors, but not both.

In fewer than four spatial dimensions, how might we devise an intrinsically stable quantum memory, analogous to a magnetic domain with long-range order that encodes a robust classical bit? Perhaps we can build a two-dimensional material with a topologically degenerate ground state, such that errors create point defects that have infinite-range attractive interactions. That system's quasi-long-range order at nonzero temperature could stabilize an arbitrary coherent superposition of ground states.

\section{Conclusions}
\label{sec:conclusions}

In foreseeable quantum computers, the quantum gates that can be executed with good fidelity are likely to be {\em local} gates --- only interactions between qubits that are close to one another will be accurately controllable. Therefore, it is important to contemplate the capabilities of large-scale quantum computers in which all gates are local in three-dimensional space. It is also reasonable to imagine that future quantum computers will include some kind of integrated classical processors, and that the classical processors will be much more accurate and much faster than the quantum processors. 

Such considerations have led us to investigate the efficacy of quantum error correction in a computational model in which all quantum gates are local, and in which classical computations of polynomial size can be done instantaneously and with perfect accuracy. We have also assumed that the measurement of a qubit can be done as quickly as the execution of a quantum gate. 

These conditions are ideally suited for the use of topological quantum error-correcting codes, such that all quantum computations needed to extract an error syndrome have excellent locality properties. Indeed, we have shown that if the two-dimensional surface codes introduced in \cite{kitaev1,kitaev2} are used, then an accuracy threshold for quantum storage can be established, and we have estimated its numerical value. This accuracy threshold can be interpreted as a critical point of a three-dimensional lattice gauge theory with quenched randomness, where the third dimension represents time. There is also an accuracy threshold for universal quantum computation, but we have not calculated it carefully.

Topological codes provide a compelling framework for controlling errors in a quantum system via local quantum processing; for this reason, we expect these codes to figure prominently in the future evolution of quantum technologies. In any case, our analysis amply illustrates that principles from statistical physics and topology can be fruitfully applied to the daunting task of accurately manipulating intricate quantum states.

\acknowledgments

We happily acknowledge helpful discussions with many colleagues, including Dorit Aharonov, Charlie Bennett, Daniel Gottesman, Randy Kamien, Greg Kuperberg, Paul McFadden, Michael Nielsen, Peter Shor, Andrew Steane, Chenyang Wang, and Nathan Wozny.  We are especially grateful to Peter Hoyer for discussions of efficient perfect matching algorithms. This work originated in 1997, while E.~D. received support from Caltech's Summer
Undergraduate Research Fellowship (SURF) program. This work has been supported in part by the Department of Energy under Grant No. DE-FG03-92-ER40701, by DARPA through the Quantum Information and Computation (QUIC) project administered by the Army Research Office under Grant No. DAAH04-96-1-0386, by the National Science Foundation under Grant No. EIA-0086038, by the Caltech MURI Center for Quantum Networks under ARO Grant No. DAAD19-00-1-0374, and by an IBM Faculty Partnership Award.

\end{document}